\documentclass[twocolumn]{aastex62}
\pdfoutput=1 
\usepackage{amsmath,amstext}
\usepackage[T1]{fontenc}
\usepackage[figure,figure*]{hypcap}
\graphicspath{{./}{figures/}}

\received{?}
\revised{?}
\accepted{?}
\submitjournal{AAS}

\shorttitle{TOI-6255}
\shortauthors{Dai et al.}

\begin{document}

\title{An Earth-sized Planet on the Verge of Tidal Disruption}

\correspondingauthor{Fei Dai}
\email{fdai@hawaii.edu}

\author[0000-0002-8958-0683]{Fei Dai}
\affiliation{Institute for Astronomy, University of Hawai`i, 2680 Woodlawn Drive, Honolulu, HI 96822, USA}
\affiliation{Division of Geological and Planetary Sciences,
1200 E California Blvd, Pasadena, CA, 91125, USA}
\affiliation{Department of Astronomy, California Institute of Technology, Pasadena, CA 91125, USA}

\author[0000-0001-8638-0320]{Andrew W. Howard}
\affiliation{Department of Astronomy, California Institute of Technology, Pasadena, CA 91125, USA}

\author[0000-0003-1312-9391]{Samuel Halverson}
\affiliation{Jet Propulsion Laboratory, California Institute of Technology, 4800 Oak Grove Drive, Pasadena, CA 91109, USA}

\author[0000-0003-2066-8959]{Jaume Orell-Miquel}
\affiliation{Instituto de Astrofísica de Canarias (IAC), Calle Vía Láctea s/n, 38205 La Laguna, Tenerife, Spain }

\author[0000-0003-0987-1593]{Enric Pall\'e}
\affiliation{Instituto de Astrofísica de Canarias (IAC), Calle Vía Láctea s/n, 38205 La Laguna, Tenerife, Spain }

\author[0000-0002-0531-1073]{Howard Isaacson}
\affiliation{{Department of Astronomy,  University of California Berkeley, Berkeley CA 94720, USA}}
\affiliation{Centre for Astrophysics, University of Southern Queensland, Toowoomba, QLD, Australia}

\author[0000-0003-3504-5316]{Benjamin Fulton}
\affiliation{NASA Exoplanet Science Institute/Caltech-IPAC, MC 314-6, 1200 E California Blvd, Pasadena, CA 91125, USA}

\author[0000-0002-3286-3543]{Ellen M. Price}
\altaffiliation{Heising-Simons Foundation 51 Pegasi b Postdoctoral Fellow}
\affiliation{Department of the Geophysical Sciences, University of Chicago, 5734 South Ellis Ave., Chicago, IL 60637, USA}

\author[0000-0002-9479-2744]{Mykhaylo Plotnykov}
\affiliation{Centre for Planetary Sciences, University of Toronto, 1265 Military Trail, Toronto, ON, M1C 1A4, Canada}

\author[0000-0003-0638-3455]{Leslie A. Rogers}
\affiliation{Department of Astronomy and Astrophysics, University of Chicago, Chicago, IL 60637, USA}

\author[0000-0003-3993-4030]{Diana Valencia}
\affiliation{Centre for Planetary Sciences, University of Toronto, 1265 Military Trail, Toronto, ON, M1C 1A4, Canada}

\author[0000-0003-0062-1168]{Kimberly Paragas}
\affiliation{Division of Geological and Planetary Sciences,
1200 E California Blvd, Pasadena, CA, 91125, USA}

\author[0000-0002-0371-1647]{Michael Greklek-McKeon}
\affiliation{Division of Geological and Planetary Sciences,
1200 E California Blvd, Pasadena, CA, 91125, USA}

\author[0000-0002-0672-9658]{Jonathan Gomez Barrientos}
\affiliation{Division of Geological and Planetary Sciences,
1200 E California Blvd, Pasadena, CA, 91125, USA}

\author[0000-0002-5375-4725]{Heather A. Knutson}
\affiliation{Division of Geological and Planetary Sciences,
1200 E California Blvd, Pasadena, CA, 91125, USA}

\author[0000-0003-0967-2893]{Erik A. Petigura}
\affiliation{Department of Physics \& Astronomy, University of California Los Angeles, Los Angeles, CA 90095, USA}

\author[0000-0002-3725-3058]{Lauren M. Weiss}
\affiliation{Department of Physics and Astronomy, University of Notre Dame, Notre Dame, IN 46556, USA}

\author[0000-0001-7058-4134]{Rena Lee}
\affiliation{Institute for Astronomy, University of Hawai`i, 2680 Woodlawn Drive, Honolulu, HI 96822, USA}

\author[0000-0002-4480-310X]{Casey L. Brinkman}
\affiliation{Institute for Astronomy, University of Hawai`i, 2680 Woodlawn Drive, Honolulu, HI 96822, USA}

\author[0000-0001-8832-4488]{Daniel Huber}
\affiliation{Institute for Astronomy, University of Hawai`i, 2680 Woodlawn Drive, Honolulu, HI 96822, USA}
\affiliation{Sydney Institute for Astronomy (SIfA), School of Physics, University of Sydney, NSW 2006, Australia}

\author[0000-0001-7409-5688]{Guðmundur Stef\'ansson}
\affiliation{Anton Pannekoek Institute for Astronomy, University of Amsterdam, Science Park 904, 1098 XH Amsterdam, The Netherlands}

\author[0000-0003-1298-9699]{Kento Masuda}
\affiliation{Department of Earth and Space Science, Osaka University, Osaka 560-0043, Japan}

\author[0000-0002-8965-3969]{Steven Giacalone}
\affiliation{Department of Astronomy, California Institute of Technology, Pasadena, CA 91125, USA}

\author[0000-0001-9352-0248]{Cicero X. Lu}
\affiliation{Gemini Observatory/NSF’s NOIRLab, 670 N. A’ohoku Place, Hilo, HI 96720, USA}
\affiliation{Science Fellow, Gemini North}

\author[0000-0002-1426-1186]{Edwin S. Kite}
\affiliation{Department of Geophysical Sciences, University of Chicago, Chicago, IL 60637, USA}

\author[0000-0003-2215-8485]{Renyu Hu}
\affiliation{Jet Propulsion Laboratory, California Institute of Technology, Pasadena, CA 91109, USA}
\affiliation{Division of Geological and Planetary Sciences, California Institute of Technology}

\author[0000-0002-5258-6846]{Eric Gaidos}
\affiliation{Department of Earth Sciences, University of Hawai'i at M\"{a}noa, 1680 East-West Road, Honolulu, HI 96822 USA}

\author[0000-0002-0659-1783]{Michael Zhang}
\affil{Department of Astronomy \& Astrophysics, University of Chicago, Chicago, IL 60637}

\author[0000-0003-3856-3143]{Ryan A. Rubenzahl}
\altaffiliation{NSF Graduate Research Fellow}
\affiliation{Department of Astronomy, California Institute of Technology, Pasadena, CA 91125, USA}

\author[0000-0002-4265-047X]{Joshua N. Winn}
\affiliation{Department of Astrophysical Sciences, Princeton University, 4 Ivy Lane, Princeton, NJ 08544, USA}

\author[0000-0002-7127-7643]{Te Han}
\affiliation{Department of Physics \& Astronomy, The University of California, Irvine, Irvine, CA 92697, USA}

\author[0000-0001-7708-2364]{Corey Beard}
\altaffiliation{NASA FINESST Fellow}
\affiliation{Department of Physics \& Astronomy, The University of California, Irvine, Irvine, CA 92697, USA}

\author[0000-0002-5034-9476]{Rae Holcomb}
\affiliation{Department of Physics \& Astronomy, University of California Irvine, Irvine, CA 92697, USA}

\author[0000-0002-5812-3236]{Aaron Householder}
\affiliation{Department of Earth, Atmospheric and Planetary Sciences, Massachusetts Institute of Technology, Cambridge, MA 02139, USA}
\affil{Kavli Institute for Astrophysics and Space Research, Massachusetts Institute of Technology, Cambridge, MA 02139, USA}

\author[0000-0003-0742-1660]{Gregory J. Gilbert}
\affiliation{Department of Physics \& Astronomy, University of California Los Angeles, Los Angeles, CA 90095, USA}

\author[0000-0001-8342-7736]{Jack Lubin}
\affiliation{Department of Physics \& Astronomy, University of California Los Angeles, Los Angeles, CA 90095, USA}

\author[0000-0001-7664-648X]{J. M. Joel Ong}
\affiliation{Institute for Astronomy, University of Hawai`i, 2680 Woodlawn Drive, Honolulu, HI 96822, USA}
\affiliation{NASA Hubble Fellow}

\author[0000-0001-7047-8681]{Alex S. Polanski}
\affil{Department of Physics and Astronomy, University of Kansas, Lawrence, KS 66045, USA}

\author[0000-0003-2657-3889]{Nicholas Saunders}
\altaffiliation{NSF Graduate Research Fellow}
\affiliation{Institute for Astronomy, University of Hawai`i, 2680 Woodlawn Drive, Honolulu, HI 96822, USA}

\author[0000-0002-4290-6826]{Judah Van Zandt}
\affil{Department of Physics \& Astronomy, University of California Los Angeles, Los Angeles, CA 90095, USA}

\author[0000-0001-7961-3907]{Samuel W.\ Yee}
\affiliation{Center for Astrophysics \textbar \ Harvard \& Smithsonian, 60 Garden Street, Cambridge, MA 02138, USA}
\altaffiliation{Heising-Simons Foundation 51 Pegasi b Postdoctoral Fellow}

\author[0000-0002-2696-2406]{Jingwen Zhang}
\altaffiliation{NASA FINESST Fellow}
\affiliation{Institute for Astronomy, University of Hawai`i, 2680 Woodlawn Drive, Honolulu, HI 96822, USA}

\author{Jon Zink}
\affiliation{Department of Astronomy, California Institute of Technology, Pasadena, CA 91125, USA}

\author[0000-0002-6153-3076]{Bradford Holden}
\affil{University of California Observatories
University of California, Santa Cruz
1156 High St, Santa Cruz, CA 95064}

\author{Ashley Baker}
\affil{Caltech Optical Observatories, Pasadena, CA, 91125, USA}

\author[0009-0008-9808-0411]{Max Brodheim}
\affiliation{W. M. Keck Observatory, 65-1120 Mamalahoa Hwy, Waimea, HI 96743}

\author[0000-0002-1835-1891]{Ian J. M. Crossfield}
\affiliation{Department of Physics \& Astronomy, University of Kansas, 1082 Malott,1251 Wescoe Hall Dr., Lawrence, KS 66045, USA}

\author[0009-0000-3624-1330]{William Deich}
\affil{University of California Observatories
University of California, Santa Cruz
1156 High St, Santa Cruz, CA 95064}

\author{Jerry Edelstein}
\affil{Space Sciences Laboratory, University of California Berkeley, Berkeley, CA 94720, USA}

\author[0009-0004-4454-6053]{Steven R. Gibson}
\affil{Caltech Optical Observatories, Pasadena, CA, 91125, USA}

\author[0000-0002-7648-9119]{Grant M. Hill}
\affiliation{W. M. Keck Observatory, 65-1120 Mamalahoa Hwy, Waimea, HI 96743}

\author[0009-0006-2900-0401]{Sharon R Jelinsky}
\affil{Space Sciences Laboratory, University of California Berkeley, Berkeley, CA 94720, USA}

\author[0000-0001-8414-8771]{Marc Kassis}
\affiliation{W. M. Keck Observatory, 65-1120 Mamalahoa Hwy, Waimea, HI 96743}

\author[0000-0003-2451-5482]{Russ R. Laher}
\affiliation{NASA Exoplanet Science Institute/Caltech-IPAC, MC 314-6, 1200 E California Blvd, Pasadena, CA 91125, USA}

\author[0009-0004-0592-1850]{Kyle Lanclos}
\affiliation{W. M. Keck Observatory, 65-1120 Mamalahoa Hwy, Waimea, HI 96743}

\author[0000-0001-7323-7277]{Scott Lilley}
\affiliation{W. M. Keck Observatory, 65-1120 Mamalahoa Hwy, Waimea, HI 96743}

\author[0009-0008-4293-0341]{Joel N. Payne}
\affiliation{W. M. Keck Observatory, 65-1120 Mamalahoa Hwy, Waimea, HI 96743}

\author{Kodi Rider}
\affil{Space Sciences Laboratory, University of California Berkeley, Berkeley, CA 94720, USA}

\author[0000-0003-0149-9678]{Paul Robertson}
\affiliation{Department of Physics \& Astronomy, University of California Irvine, Irvine, CA 92697, USA}

\author[0000-0001-8127-5775]{Arpita Roy}
\affiliation{Astrophysics \& Space Institute, Schmidt Sciences, New York, NY 10011, USA}

\author[0000-0002-4046-987X]{Christian Schwab}
\affiliation{School of Mathematical and Physical Sciences, Macquarie University, Balaclava Road, North Ryde, NSW 2109, Australia}

\author[0000-0003-3133-6837]{Abby P. Shaum}
\affiliation{Department of Astronomy, California Institute of Technology, Pasadena, CA 91125, USA}

\author[0009-0007-8555-8060]{Martin M. Sirk}
\affil{Space Sciences Laboratory, University of California Berkeley, Berkeley, CA 94720, USA}

\author{Chris Smith}
\affil{Space Sciences Laboratory, University of California Berkeley, Berkeley, CA 94720, USA}

\author[0009-0003-6534-0428]{Adam Vandenberg}
\affiliation{W. M. Keck Observatory, 65-1120 Mamalahoa Hwy, Waimea, HI 96743}

\author[0000-0002-6092-8295]{Josh Walawender}
\affiliation{W. M. Keck Observatory, 65-1120 Mamalahoa Hwy, Waimea, HI 96743}

\author[0000-0002-6937-9034]{Sharon X.~Wang}
\affiliation{Department of Astronomy, Tsinghua University, Beijing 100084, People's Republic of China}

\author[0009-0002-4850-5377]{Shin-Ywan (Cindy) Wang}
\affiliation{NASA Exoplanet Science Institute/Caltech-IPAC, MC 314-6, 1200 E California Blvd, Pasadena, CA 91125, USA}

\author{Edward Wishnow}
\affil{Space Sciences Laboratory, University of California Berkeley, Berkeley, CA 94720, USA}

\author[0000-0001-6160-5888]{Jason T.\ Wright}
\affiliation{Department of Astronomy \& Astrophysics, 525 Davey Laboratory, Penn State, University Park, PA, 16802, USA}
\affiliation{Center for Exoplanets and Habitable Worlds, 525 Davey Laboratory, Penn State, University Park, PA, 16802, USA}
\affiliation{Penn State Extraterrestrial Intelligence Center, 525 Davey Laboratory, Penn State, University Park, PA, 16802, USA}

\author[0000-0002-4037-3114]{Sherry Yeh}
\affiliation{W. M. Keck Observatory, 65-1120 Mamalahoa Hwy, Waimea, HI 96743}


\author[0000-0002-7349-1387]{Jos\'e~A.~Caballero} 
\affiliation{Centro de Astrobiolog\'ia (CSIC-INTA), ESAC campus, Camino Bajo del Castillo s/n, 28692 Villanueva de la Ca\~nada, Madrid, Spain}

\author[0000-0003-0061-518X]{Juan C. Morales} 
\affiliation{Institut de Ci\`encies de l’Espai (ICE, CSIC), Campus UAB, Can Magrans s\/n, 08193 Bellaterra, Barcelona, Spain }
\affiliation{Institut d’Estudis Espacials de Catalunya (IEEC), 08034 Barcelona, Spain}

\author[0000-0001-9087-1245]{Felipe Murgas}
\affiliation{Instituto de Astrofísica de Canarias (IAC), Calle Vía Láctea s/n, 38205 La Laguna, Tenerife, Spain }

\affiliation{Departamento de Astrof\'isica, Universidad de La Laguna (ULL), 38206 La Laguna, Tenerife, Spain}

\author[0000-0002-4019-3631]{Evangelos Nagel}
\affiliation{Institut f\"ur Astrophysik und Geophysik, Georg-August-Universit\"at  G\"ottingen, Friedrich-Hund-Platz 1, 37077 G\"ottingen, Germany}

\author[0000-0003-1242-5922]{Ansgar Reiners} 
\affiliation{Institut f\"ur Astrophysik und Geophysik, Georg-August-Universit\"at  G\"ottingen, Friedrich-Hund-Platz 1, 37077 G\"ottingen, Germany}

\author[0000-0002-1624-0389]{Andreas Schweitzer}
\affiliation{Hamburger Sternwarte, Gojenbergsweg 112, D-21029 Hamburg, Germany}

\author[0000-0002-8087-4298]{Hugo M. Tabernero}
\affiliation{Departamento de F{\'i}sica de la Tierra y Astrof{\'i}sica \& IPARCOS-UCM (Instituto de F\'{i}sica de Part\'{i}culas y del Cosmos de la UCM), Facultad de Ciencias F{\'i}sicas, Universidad Complutense de Madrid, 28040 Madrid, Spain}

\author[0000-0002-6532-4378]{Mathias Zechmeister} 
\affiliation{Institut f\"ur Astrophysik und Geophysik, Georg-August-Universit\"at  G\"ottingen, Friedrich-Hund-Platz 1, 37077 G\"ottingen, Germany}


\author[0000-0001-9263-6775]{Alton Spencer}
\affiliation{Western Connecticut State University, Danbury, CT, 06810}

\author[0000-0002-5741-3047]{David R. Ciardi}
\affiliation{NASA Exoplanet Science Institute/Caltech-IPAC, 1200 E. California Blvd., Pasadena, CA 91125, USA}

\author[0000-0002-2361-5812]{Catherine A. Clark}
\affiliation{Jet Propulsion Laboratory, California Institute of Technology, 4800 Oak Grove Drive, Pasadena, CA 91109, USA}
\affiliation{NASA Exoplanet Science Institute/Caltech-IPAC, 1200 E. California Blvd., Pasadena, CA 91125, USA}

\author[0000-0003-2527-1598]{Michael B. Lund}
\affiliation{NASA Exoplanet Science Institute/Caltech-IPAC, 1200 E. California Blvd., Pasadena, CA 91125, USA}

\author[0000-0003-1963-9616]{Douglas A. Caldwell}
\affiliation{SETI Institute, Mountain View, CA 94043 USA/NASA Ames Research Center, Moffett Field, CA 94035 USA}

\author[0000-0001-6588-9574]{Karen A.\ Collins}
\affiliation{Center for Astrophysics \textbar \ Harvard \& Smithsonian, 60 Garden Street, Cambridge, MA 02138, USA}

\author[0000-0001-8227-1020]{Richard P. Schwarz}
\affiliation{Center for Astrophysics \textbar \ Harvard \& Smithsonian, 60 Garden Street, Cambridge, MA 02138, USA}

\author[0000-0003-1464-9276]{Khalid Barkaoui}
\affiliation{Astrobiology Research Unit, Universit\'e de Li\`ege, 19C All\'ee du 6 Ao\^ut, 4000 Li\`ege, Belgium}
\affiliation{Department of Earth, Atmospheric and Planetary Science, Massachusetts Institute of Technology, 77 Massachusetts Avenue, Cambridge, MA 02139, USA}
\affiliation{Instituto de Astrof\'isica de Canarias (IAC), Calle V\'ia L\'actea s/n, 38200, La Laguna, Tenerife, Spain}

\author[0000-0001-8621-6731]{Cristilyn N.\ Watkins}
\affiliation{Center for Astrophysics \textbar \ Harvard \& Smithsonian, 60 Garden Street, Cambridge, MA 02138, USA}

\author[0000-0002-1836-3120]{Avi Shporer}
\affiliation{Department of Physics and Kavli Institute for Astrophysics and Space Research, Massachusetts Institute of Technology, Cambridge, MA 02139, USA}

\author[0000-0001-8511-2981]{Norio Narita}
\affiliation{Komaba Institute for Science, The University of Tokyo, 3-8-1 Komaba, Meguro, Tokyo 153-8902, Japan}
\affiliation{Astrobiology Center, 2-21-1 Osawa, Mitaka, Tokyo 181-8588, Japan}
\affiliation{Instituto de Astrofísica de Canarias (IAC), Calle Vía Láctea s/n, 38205 La Laguna, Tenerife, Spain }

\author[0000-0002-4909-5763]{Akihiko Fukui}
\affiliation{Komaba Institute for Science, The University of Tokyo, 3-8-1 Komaba, Meguro, Tokyo 153-8902, Japan}
\affiliation{Instituto de Astrofísica de Canarias (IAC), Calle Vía Láctea s/n, 38205 La Laguna, Tenerife, Spain }
\author{Gregor Srdoc}
\affil{Kotizarovci Observatory, Sarsoni 90, 51216 Viskovo, Croatia}

\author[0000-0001-9911-7388]{David W. Latham}
\affiliation{Center for Astrophysics \textbar Harvard \& Smithsonian, 60 Garden St, Cambridge, MA 02138, USA}

\author[0000-0002-4715-9460]{Jon M. Jenkins}
\affiliation{NASA Ames Research Center, Moffett Field, CA 94035, USA}

\author[0000-0003-2058-6662]{George R. Ricker}
\affiliation{Department of Physics and Kavli Institute for Astrophysics and Space Research, Massachusetts Institute of Technology, Cambridge, MA 02139, USA}

\author[0000-0002-6892-6948]{Sara Seager}
\affiliation{Department of Physics and Kavli Institute for Astrophysics and Space Research, Massachusetts Institute of Technology, Cambridge, MA
02139, USA}
\affiliation{Department of Earth, Atmospheric and Planetary Sciences, Massachusetts Institute of Technology, Cambridge, MA 02139, USA}
\affiliation{Department of Aeronautics and Astronautics, MIT, 77 Massachusetts Avenue, Cambridge, MA 02139, USA}

\author[0000-0001-6763-6562]{Roland Vanderspek}
\affiliation{Department of Physics and Kavli Institute for Astrophysics and Space Research, Massachusetts Institute of Technology, Cambridge, MA 02139, USA}



\begin{abstract}
\noindent TOI-6255~b (GJ 4256) is an Earth-sized planet (1.079$\pm0.065$ $R_\oplus$) with an orbital period of only 5.7 hours. With the newly commissioned Keck Planet Finder (KPF) and CARMENES spectrographs, we determined the planet's mass to be 1.44$\pm$0.14 $M_{\oplus}$. The planet is just outside the Roche limit, with $P_{\rm orb}/P_{\rm Roche}$ = 1.13 $\pm0.10$.  The strong tidal force likely deforms the planet into a triaxial ellipsoid with a long axis that is $\sim$10\% longer than the short axis.  Assuming a reduced stellar tidal quality factor $Q_\star^\prime \approx10^7$, we predict that tidal orbital decay will cause TOI-6255 to reach the Roche limit in roughly 400 Myr. Such tidal disruptions may produce the possible signatures of planet engulfment that have been on stars with anomalously high refractory elemental abundances compared to its conatal binary companion. TOI-6255 b is also a favorable target for searching for star-planet magnetic interactions, which might cause interior melting and hasten orbital decay. TOI-6255 b is a top target (Emission Spectroscopy Metric of about 24) for phase
curve observations with the James Webb Space Telescope. 

\end{abstract}


\section{Introduction}
The ``ultra-short-period'' (USP) planets are generally terrestrial planets ($R<2\,R_\oplus$) that orbit their host stars in less than one day \citep{USP}. More than a hundred USPs have been described in the literature. Their occurrence rate around sun-like stars is about 0.5\% \citep{USP}. USPs provide favorable opportunities to study the composition of terrestrial planets \citep{Winn2018}. A true Earth analog on a yearly orbit induces a Doppler wobble of 9~cm s$^{-1}$, which is beyond the capabilities of current-generation spectrographs. The same planet on a sub-day orbit would induce a Doppler wobble of a few m s$^{-1}$, which is within the reach of current spectrographs. Moreover, the irradiation of a USP by its host star is more than a thousand times the Earth's insolation. There is both theoretical and empirical evidence that any primordial H/He atmosphere on these planets would have been eroded by intensive mass loss \citep{USP,Lundkvist,Lopez2017}. Thus, one can directly make use of the mass and radius
measurements to probe the planet's interior composition. USP planets are also amenable to observations of thermal phase curves and secondary eclipses with the James Webb Space Telescope (JWST). The albedo, phase offset, day-night temperature contrast, and emission spectra might reveal the presence or lack of an outgassed secondary atmosphere. In the absence of an atmosphere, phase curve observation may reveal the dominant surface mineralogy on these rocky planets \citep{Hu2012,Demory,LHS3844,Whittaker,Zhang2024}.

Several recent works suggested that the photospheric composition of a star can be significantly altered by the engulfment of rocky planets and their refractory elements \citep{Ram,Oh,Nagar,Galarza,Behmard,Liu2024}. Specifically, there are examples of stars for which the refractory elemental abundances are enhanced by about 0.1 dex relative to a co-moving and presumably co-natal binary companion. The USPs are the natural candidates for the planets that are engulfed and produce such signatures. Tidal interactions with the host star gradually shrink the orbit of a USP, and eventually bring them within the tidal disruption limit \citep{Rappaport}. The resulting debris would fall onto the surface convective layer of the host star and alter the photospheric abundances. The strength of tidal interaction is steeply dependent on the orbital separation (Eqn \ref{eqn:tidal_decay}); thus, only the shortest-period planets may be tidally disrupted or engulfed by the host star within the age of the universe \citep{Winn2018,Jia}. In this work, we present the discovery of perhaps the most compelling planet that is doomed for engulfment. TOI-6255 b is an Earth-sized planet with an extraordinarily short orbital period of merely 5.7 hours. Tidal disruption could happen in the next 400\,Myr if the host star has a reduced tidal quality factor $Q_\star^\prime=10^7$.

In Section 2, we characterize the host star TOI-6255 in detail. We measure the radius of TOI-6255 b in Section 3 using transits. We extract the mass of the planet from Doppler observations in Section 4. In Section 5, we discuss the composition, tidal decay, tidal distortion, and phase curve of TOI-6255 b, as well as the suspected star-planet magnetic interaction. Section 6 is a summary of the paper.

\section{Host Star Properties}\label{sec:stellar_para}
\subsection{Spectroscopic Analysis}

We obtained a high-resolution, high-signal-to-noise-ratio (SNR), iodine-free spectrum of TOI-6255 
with the High Resolution
Echelle Spectrometer on the 10-meter Keck Telescope \citep[Keck/HIRES;][]{Vogt}. The observation was taken on UT Nov 26th, 2023. We exposed for 900 seconds and reached a signal-to-noise ratio (SNR) of about 80 per reduced pixel near 550\,nm.  We utilized the {\tt SpecMatch-Emp} pipeline \citep{Yee2017} to extract the spectroscopic parameters ($T_{\rm eff}$ and [Fe/H]). In short, {\tt SpecMatch-Emp} cross-matches the observed spectrum of a star with a library of hundreds of well-calibrated stellar spectra observed by the California Planet Search collaboration. This empirical approach circumvents many known systematic effects that plague the direct spectral modeling of low-mass stars. Some of these systematic effects include the poor definition of a continuum and the imperfect molecular line list. {\tt SpecMatch-Emp} gives an effective temperature $T_{\rm eff}=3421\pm70$~K and a metallicity [Fe/H] = --0.14$\pm$0.09.  

We then included the Gaia parallax
\citep[49.0544$\pm$0.0236 mas,][]{gaiaDR3} in our analysis. One can obtain a direct constraint on the stellar radius using the Stefan-Boltzmann law. The effective temperature, the $K$-band magnitude (minimal extinction), and the parallax (distance) of a star together constrain its radius. We used the Python package {\tt Isoclassify} \citep{Huber} to carry out this calculation. We adopted the MESA Isochrones \& Stellar Tracks \citep[MIST,][]{MIST}. We used default settings recommended by {\tt Isoclassify}. Tab. 1 summarizes the posterior distribution of the stellar parameters of TOI-6255. We note that {\tt Isoclassify} results do not include any systematic errors that may be present in different theoretical model grids. \citet{Tayar} estimated a $\sim2\%$ error on $T_{\rm eff}$, $\sim4\%$ on $M_{\star}$, and $\sim5\%$ on $R_{\star}$.

To check on the systematic errors in stellar parameters, we performed an independent analysis using the CARMENES template spectrum of TOI-6255 (see Section \ref{sec:carmenes} for details of the CARMENES observations). In short, the CARMENES data were analyzed with the publicly available {\tt
SteParSyn} package \citep{Tabernero} in combination with the procedures described in \citet{Schweitzer2019} and \citet{Cifuentes2020}. While,  {\tt Isoclassify} uses the MIST models, the {\tt
SteParSyn} parameter determination used a synthetic model grid computed using BT-Settl \citep{Allard} models as described in \citep{Marfil}.  {\tt Isoclassify} and {\tt
SteParSyn} results agree well within 1$\sigma$: $T_{\rm eff} = 3421\pm70$\,K ({\tt Isoclassify}) vs. $3455\pm70$\,K ({\tt SteParSyn}); $\log{g} = 4.850\pm0.044$  vs. $4.78\pm0.05$; [Fe/H]$ = -0.14\pm0.09$  vs. $-0.16\pm0.06$; $M_\star = 0.353\pm0.015\,M_\odot$  vs. $0.357\pm0.019\,M_\odot$; $R_\star  = 0.370\pm0.011\,R_\odot$  vs. $0.361\pm0.015\,R_\odot$. 
All these values are also consistent with previous determinations by 
\citet{Lepine2013}
and \citet{Gaidos2014}.

We also checked whether there is infrared excess due to the presence of a debris disk using the stellar photosphere models with stellar parameters derived from the CARMENES spectrum of TOI-6255. We plotted 2MASS $JHK_s$ bands, WISE $W1$, $W2$, $W3$, and $W4$ photometry, and found no evidence for infrared excess out to $22\,\mu$m against a BT-Settl model, in agreement with the results of \citet{Cifuentes}. However, the WISE bands are not sensitive to colder debris disks.
Finally, with the $K_s$ magnitude and the Gaia parallax as input, the widely used empirical relation by \citet{Mann2019} also gives a mass of $0.3591\pm0.0086 M_\odot$, which is consistent with the {\tt Isoclassify} within 1$\sigma$. We chose to use the {\tt Isoclassify} results for further analysis.

\subsection{Nearby Stellar Companion}
We checked if TOI-6255 has a nearby stellar companion that may produce a false positive transit signal. We observed TOI-6255 with Palomar/PHARO \citep{hayward2001} on UT Jun 30th, 2023.  No stellar companion was detected in the narrow-band Br$\gamma$ filter $(\lambda_o$ = 2.1686\,$\mu$m; $\Delta\lambda$ = 0.0326\,$\mu$m). Each dither position was observed three times, offset in position from each other by 0.5\arcsec\ for a total of 15 frames; with an integration time of 5.7 seconds per frame, the total on-source time was 85.5 seconds. Fig. \ref{fig:palomar} shows the resultant contrast limit as a function of angular separation. 

Using Gaia DR3 data \citep{gaiaDR3}, no comoving companion was identified within a radius of 10 arcmin. The Gaia Renormalised Unit Weight Error (RUWE) can be considered as a reduced $\chi^2$ of their single-star astrometric solution. TOI-6255 has a RUWE of 1.39 just below the threshold of 1.4 below which the astrometric solution is considered consistent with a single star. 
This is consistent with the previous high-resolution imaging survey for close companions performed by \citet{Contreras}.

The proper motion of TOI-6255 reported by Gaia does not match with any known comoving associations reported in {\tt Banyan-$\Sigma$} \citep{Gagne} and in \citet{Bouma_2022}. Using the framework of \citet{Bensby}, TOI-6255 has about 4\% chance of being in the thick disk based on its Galactic UVW velocity (U, V, W = 32.0$\pm1.0$, -15.0$\pm2.1$, -27.2$\pm0.5$ km~s$^{-1}$) after correcting for the Local Standard of Rest. In this calculation, we used the values: U$_\odot$,V$_\odot$,W$_\odot$ = 10.0$\pm1.0$, 11.0$\pm2.0$, 7.0$\pm0.5$ km~s$^{-1}$ from \citet{Bland-Hawthorn}.

\subsection{Host Star Age}

While many methods of estimating stellar ages do not apply to M dwarfs, gyrochronology (relating declining rotation rate to increasing age through temperature-dependent relations) may still be applicable.  No periodic signal is obvious in either a Lomb-Scargle periodogram \citep{Lomb1976,Scargle1982} or an auto-correlation analysis \citep{McQuillan2014} of \emph{TESS} lightcurves of TOI-6255, probably because the rotation periods of middle-aged mid-type M dwarfs are longer than the \emph{TESS} orbit (and lightcurve systematics timescale) of 13.7 days.  Each {\it TESS} sector is 27-days in duration, and only two sectors have been observed for TOI-6255. Therefore, measuring the stellar rotation of TOI-6255 using {\it TESS} data can be challenging.  A periodogram of a lightcurve from the longer-duration WASP survey \citep{WASP} contains two peaks at $\sim$68 and 85 days: application of the \citet{Gaidos2023} gyrochronology to the former yields an age of $6 \pm 2$ Gyr, consistent with main sequence status.  However, we caution that this gyrochronology is poorly calibrated at the cool end, and stars like TOI-6255 near or at the fully convective boundary can have markedly different spin-down rates \citep[e.g.,][]{Chiti}.

\begin{deluxetable*}{lcc}
\tablecaption{Stellar Parameters of TOI-6255 \citep[GJ~4256][]{Gliese}} 
\label{tab:stellar_para}
\tablehead{
\colhead{Parameters} & \colhead{Value and 68.3\% Confidence Interval} & \colhead{Reference}}
\startdata
TIC ID  & 261135533 & A\\
R.A.  & 22:06:00.76 & A\\
Dec.  & +39:17:55.8  & A\\ 
$V$ (mag) & 12.747 $\pm$ 0.052& A\\
$K_s$ (mag) & 8.071 $\pm$ 0.02& A\\
Effective Temperature $T_{\text{eff}} ~(K)$ & $3421\pm70$ & B \\
Surface Gravity $\log~g~(\text{cm~s}^{-2})$ &$4.850 \pm 0.044$& B \\
Iron Abundance $[\text{Fe/H}]~(\text{dex})$ &$-0.14 \pm 0.09$& B \\
Stellar Mass $M_{\star} ~(M_{\odot})$ &$0.353\pm0.015$& B \\
Stellar Radius $R_{\star} ~(R_{\odot})$ &$0.370\pm0.011$& B \\
Stellar Density $\rho_\star$ (g cm$^{-3}$) &$9.8\pm1.0$&B \\
Limb Darkening q$_1$ \citep{Kipping} & $0.50\pm0.28$& B\\
Limb Darkening q$_2$ \citep{Kipping} & $0.35\pm0.18$& B\\
Parallax $\pi$ (mas)&$49.0544\pm0.0236$& C \\
\enddata
\tablecomments{A:TICv8; B: this work; C: \citet{Gaia}}
\end{deluxetable*}

\section{Photometric Observations}
\label{sec:photometry}
\subsection{TESS Observations}
{\it TESS} \citep{Ricker} observed TOI-6255 during Sectors 16 and 56 in September 2019 and in September 2022. We downloaded the 2-min cadence light curve produced by the {\it TESS} Science Processing Operations Center \citep[SPOC of NASA Ames Research Center,][]{jenkinsSPOC2016}. The data is available on the Mikulski Archive for Space
Telescopes website\footnote{\url{https://archive.stsci.edu}} or the following \dataset[DOI]{doi:10.17909/t9-nmc8-f686}\citep{doi}. Our analysis made use of both the Presearch Data Conditioning Simple Aperture Photometry \citep[PDC-SAP;][]{Stumpe2014} light curve and the Simple Aperture Photometry \citep[SAP, ][]{Twicken2010,Morris2020}. PDC-SAP light curve underwent more extensive systematic mitigation and is hence used for modeling the transit signal. On the other hand, SAP light curve better preserves any long-term stellar variability, it was used in our phase curve analysis and the measurement of the rotation period for the host star. We discarded all data points that had a non-zero Quality Flag.

\subsection{Additional Transiting Planets?}
The TESS Science Processing Operations Center \citep[SPOC at NASA Ames Research Center, ][]{Jenkins2016SPOC} initially detected the transit of TOI-6255.01 (0.238-day planet). We searched for other transiting signals in the light curve. We first removed any stellar activity or instrumental variation by fitting a cubic spline in time of a width of 0.75 days. We searched the detrended light curve with our own Box-Least-Square code \citep[BLS,][]{Kovac2002} that was used in previous works \citep[e.g.][]{Dai_1444}. We recovered TOI-6255.01, as well as another 14.48-day transit-like signal TOI-6255.02 with a signal detection efficiency \citep[defined by][]{Ofir2014} of 14.3. This signal was independently reported on the ExoFOP website \footnote{\url{https://exofop.ipac.caltech.edu/tess/}} by citizen scientist Alton Spencer. SPOC analysis of TOI-6255.02 showed a substantial flux centroid shift of 27$\pm3$ arcsecond towards a nearby background star. This suggests that TOI-6255.02 is most likely a false positive. We are unable to detect any other prominent transit signal from the current data set. 

\subsection{Transit Modeling}\label{sec:transit_model}

Our transit model is based on the {\tt Python} package {\tt Batman} \citep{Kreidberg2015}. The stellar density of the host star is a parameter for which a prior can be set by our spectroscopic analysis $\rho = 9.8\pm1.0$~g~cm$^{-3}$. The stellar density is crucial for breaking the degeneracy between the semi-major axis and impact parameter of the planet \citep{Seager}. For limb darkening coefficients, we adopted the formulation of \citet{Kipping} ($q_1$ and $q_2$). The transit parameters further include the orbital period $P_{\text{orb}}$, the time of conjunction $T_{\text{c}}$, the planet-to-star radius ratio $R_{\text{p}}/R_\star$, the scaled orbital distance $a/R_\star$, the cosine of the orbital inclination cos$i$, the orbital eccentricity $e$, and the argument of pericenter $\omega$. $e$ and $\omega$ are recombined to $\sqrt{e}$~cos$\omega$ and $\sqrt{e}$~sin$\omega$ to give rise to a uniform prior on eccentricity \citep{Lucy}. We initially allowed for non-zero eccentricities. However, the resultant constraint on eccentricities using the current dataset is weak (even after incorporating the radial velocity data in Section \ref{sec:rv_model}). To reduce model complexity, we adopted circular orbits.

We first fitted all transits observed by {\it TESS} assuming a linear ephemeris (i.e. constant period, no transit timing variations). The best-fit constant-period model was found by maximizing the likelihood with the {\tt Levenberg-Marquardt} method implemented in {\tt Python} package {\tt lmfit} \citep{LM}. The best-fit constant-period model then served as a template model. Each individual transit was fitted with this template model. The only free parameters are the mid-transit times and three parameters for a quadratic function of time for any out-of-transit flux variation. This process yielded a list of individual transit times. We were not able to identify any statistically meaningful transit timing variations. To sample the posterior distribution of the transit parameters, we performed a Markov Chain Monte Carlo analysis using the {\tt emcee} package \citep{emcee}. We initialized 128 walkers in the vicinity of the maximum-likelihood model from  {\tt lmfit}. We ran {\tt emcee} for 50000 steps which is much longer than the autocorrelation length of each parameter (hundreds of steps). We summarize the posterior distribution in Tab. 2. Fig. \ref{fig:transit} is the phase-folded and binned transit of TOI-6255 b. TOI-6255 b has been confirmed by our strong radial velocity (RV) detection and the well-modeled transit light curve (Section \ref{sec:rv_model}). 
 Appendix \ref{sec:app_phase_curve} shows a tentative detection of phase curve variation of TOI-6255 b in {\it TESS} light curves. Appendix \ref{sec:transits} shows additional transit modeling results using ground-based light curves. \

\begin{deluxetable*}{lll}
\tablecaption{Planetary Parameters of TOI-6255} 
\label{tab:planet_para}
\tablehead{
\colhead{Parameter}  & \colhead{Symbol} &  \colhead{Posterior Distribution} }
\startdata
TOI-6255 b\\
Planet/Star Radius Ratio & $R_p/R_\star$  & $0.0267\pm0.0014$  \\
Time of Conjunction (BJD-2457000) & $T_c$  & $1738.71248\pm0.00043$  \\
Impact Parameter & $b$  & $0.84\pm0.05$ \\
Scaled Semi-major Axis & $a/R_\star$  & $3.10\pm 0.12$  \\
Orbital Inclination (deg) & $i$  & $74.4\pm1.2$  \\
Orbital Eccentricity  & $e$  & 0 (fixed)  \\
Orbital Period (days) & $P_{\rm orb}$   & $0.23818244\pm0.00000012$  \\
RV Semi-amplitude (m/s)  & $K$ &$2.98\pm0.28$ \\
Planetary Radius ($R_\oplus$)  & $R_{\rm p}$ &1.079$\pm0.065$ \\
Planetary Mass ($M_\oplus$)  & $M_{\rm p}$ &1.44$\pm0.14$ \\
\hline
KPF RV Jitter (m~s$^{-1}$)  & $\sigma_{\rm KPF}$ &$0.7 \pm 0.3$\\
CARMENES RV Jitter (m~s$^{-1}$)  & $\sigma_{\rm CARMENES}$ &$<0.4$ (95\% confidence)\\
Kernel Amplitude  (m~s$^{-1}$) & $h$ &$8.9 \pm 1.1$\\
Correlation Timescale (days) & $\tau$ &$1.2_{-0.1}^{+0.4}$\\
\enddata
\end{deluxetable*}

\section{Spectroscopic Observations}

\subsection{KPF Observations}\label{sec:kpf_obs}
The Keck Planet Finder \citep[KPF, PI: A. Howard, ][]{Gibson2016,Gibson,Gibson2020} is an echelle spectrometer that was commissioned at the Keck Observatory in March 2023.  KPF covers the wavelength range of 445--870 nm with a resolving power of 98{,}000. Moreover, KPF's unique combination of high efficiency, 10-m aperture, and scheduling flexibility makes it well suited for studying ultra-short period planets. This paper is one of the first science results based on KPF data. 

We observed TOI-6255 with KPF from May 22nd to Nov 24 of 2023 for a total of 91 exposures. The exposure time was 10 min. We achieved a typical SNR of 87 at 550 nm.  The spectra were reduced with the KPF Data Reduction Pipeline (DRP) which is publicly available on Github\footnote{\url{https://github.com/Keck-DataReductionPipelines/KPF-Pipeline}}. The KPF DRP performs quadrant stitching, flat-fielding, order tracing, and optimal extraction. The host star is an M dwarf for which traditional cross-correlation function RV extraction is sometimes inferior to template-matching methods. The cross-correlation method is limited by the lack of a good line list for the many molecular absorption features that characterize the spectra of low-mass stars.
The continuum level and a list of isolated lines are also hard to define. We used the publicly available template-matching code \texttt{serval} \citep{SERVAL} pipeline to extract the RVs from our KPF spectra.

KPF's wavelength calibration sources include Th-Ar and U-Ne lamps, a Laser Frequency Comb, a Fabry-P\'erot Etalon, and the Solar Calibrator \citep{KPFSoCal}. Wavelength calibration sources and the wavelength calibration algorithm in the KPF DRP are still subject to fine-tuning at the time of writing this paper. Some of the apparent radial velocity variations in our data for TOI-6255 are due to night-to-night instrumental drifts. Whenever possible, we tried to obtain several exposures of the star per calendar night such that night-to-night instrumental drifts can be successfully mitigated either with a correlated noise model or a floating chunk offset technique (see Section \ref{sec:rv_model}). Such an observational strategy has proven useful in the RV characterization of many USP planets \citep[e.g.][]{Howard78}. We implemented this observation strategy using the KPF-Community Cadence algorithm described in \citet{Handley}. The KPF RVs are shown in Tab. 3 and 4.

\subsection{CARMENES Observations}\label{sec:carmenes}

TOI-6255 was also observed with CARMENES (Calar Alto high-Resolution search for M dwarfs with Exoearths with Near-infrared and optical \'Echelle Spectrographs; \citealp{Quirrenbach_2014, Quirrenbach_2020}), located at the 3.5-m telescope of the Calar Alto Observatory, Almer\'ia, Spain. CARMENES is a dual-channel spectrograph: the optical channel (VIS) covers the wavelength range from 0.52 to 0.96\,$\mu$m ($\mathcal{R}$\,=\,94,600), and the near-infrared channel (NIR) covers from 0.96 to 1.71\,$\mu$m ($\mathcal{R}$\,=\,80,400).

We obtained 33 high-resolution spectra between Jun 24th and Aug 8th, 2023. The exposure time was 30\,min, and the typical median SNR achieved in the visible channel was about 68. The spectra were reduced using the \texttt{caracal} \citep{Caballero2016} pipeline.
We corrected the spectra for telluric absorption (\citealp{Nagel2022_TAC}) and derived relative RVs with the \texttt{serval} pipeline \citep{SERVAL}. The RVs were further adjusted using measured nightly zero point corrections (\citealp{Trifonov2020_nzp}). The CARMENES RVs are shown in Tab. 5. 

\subsection{Gaussian Process Model}\label{sec:rv_model}

The measured RV variations of TOI-6255 (Fig. \ref{fig:rv})  display a correlated noise component that does not phase up with the orbital period of TOI-6255 b. As mentioned in Section \ref{sec:kpf_obs}, the correlated noise could be due to a combination of stellar activity and instrumental drifts. We employed a Gaussian Process (GP) model \citep[e.g. ][]{Haywood,Grunblatt2015} to disentangle the planetary signals from the correlated noise component.

We used the same GP framework as in our previous works \citep{Dai2019,Dai_1444}. However, instead of using a quasi-periodic kernel which is widely used in exoplanet radial velocity analyses \citep[e.g.][]{Haywood,Grunblatt2015}, we used a simpler squared-exponential kernel. In other words, we dropped the periodic component of the quasi-periodic kernel. The quasi-periodic kernel is designed to model the quasi-periodic variation of stellar activity produced by the host star rotation coupled with its surface magnetic features. For TOI-6255, we were not able to robustly identify the quasi-periodic rotational modulation of TOI-6255 with {\it TESS} observations (Section \ref{sec:photometry}). Moreover, we suspect at least some of the correlated noise is due to instrumental drift which is unrelated to the stellar rotation period. Using the squared-exponential kernel, we are agnostic to the source of the radial velocity jitter. Our covariance matrix takes the following form:

\begin{equation}
\label{covar}
C_{i,j} = h^2 \exp{\left[-\frac{(t_i-t_j)^2}{2\tau^2}\right]}+\left[\sigma_i^2+\sigma_{\text{jit}}^2\right]\delta_{i,j}
\end{equation}

where $C_{i,j}$ is the covariance matrix.
$h$ is the amplitude of the squared-exponential kernel; $\tau$ is the correlation timescale. $t_i$ is the time of each observation. $\delta_{i,j}$ is the Kronecker delta function i.e. the white noise component.  $\sigma_i$ are the nominal uncertainty produced by the RV pipelines and $\sigma_\mathrm{jit}$ is a jitter term in case we have underestimated the white noise component. With this covariance matrix, our likelihood function is:

\begin{equation}
\label{likelihood}
\log{\mathcal{L}} =  -\frac{N}{2}\log{2\pi}-\frac{1}{2}\log{|\bf{C}|}-\frac{1}{2}\bf{r}^{\text{T}}\bf{C} ^{-\text{1}} \bf{r}
\end{equation}
where $\mathcal{L}$ is the likelihood function; $N$ is the total number of RV measurements. $\bf{r}$ is the residual vector: ${\bf r}\equiv RV(t_i)-M(t_i)$ measured RV minus the model RV. The RV models are circular Keplerian described by the RV semi-amplitude $K$, the orbital period $P_{\rm orb}$, and the time of conjunction $T_c$. We imposed Gaussian priors on  $P_{\rm orb}$ and $T_c$ using the results of the transit modeling in Section \ref{sec:transit_model}.

We allowed all of the hyperparameters of the GP to float freely in our modeling of the combined KPF and CARMENES RV dataset. The only exception is that we placed an ad hoc lower boundary on the correlation timescale $\tau$ $>1$ day so that the GP does not subsume the planetary signal of TOI-6255 b on 5.7 hours. We again sampled the posterior distribution using an MCMC analysis with {\tt emcee}. The procedure is similar to that presented in Section \ref{sec:transit_model}. The results are summarized in Tab. 2. The RV variation of TOI-6255 b is robustly detected with $K=2.98\pm0.28$m~s$^{-1}$.

For readers who are concerned that a GP model may produce a spurious detection of planetary signals, we highlight the KPF observations taken on Nov 24 2023 (see the inset panel of Fig. \ref{fig:rv}). The continuous set of 21 exposures obtained that night show the expected radial-velocity variation over the 5.7-hour period of TOI-6255 b. The correlated noise component modeled by GP (blue dotted line) is essentially constant on such a short timescale. We are unable to detect another RV planet in the presence of the correlated noise.

\subsection{Floating Chunk Offset Method}\label{sec:fco}
Another commonly used strategy for dealing with correlated noise in RV datasets is the Floating Chunk Offset method \citep[e.g.][]{Hatzes2011}. If the planet of interest has an orbital period of less than 1 day, one can rely exclusively on radial-velocity variations observed during a single night, and discard any information about radial-velocity variations between nights. 
In essence, any long-term stellar activity or instrumental effects are removed by allowing each night to have an independent additive RV offset.

This method naturally requires multiple observations taken within a single night. We trimmed our RV datasets by removing the isolated exposures. We were left with 106 data points taken on 29 different calendar nights, each given an offset $\gamma_1$ to $\gamma_{29}$. In this model, the only major change to the covariance function (Eqn. \ref{covar}) is that we have dropped the squared-exponential kernel and only retained the white noise component. The likelihood is unchanged other than the implicit inclusion of the nightly offsets in $RV(t_i)$. 

We again used {\tt emcee} to sample the posterior distribution following a similar procedure as in the previous section. The Floating Chunk Offset method was able to robustly detect TOI-6255 b as well, yielding the RV semi-amplitude $K=2.93\pm0.38$m~s$^{-1}$. This is fully consistent with the value determined in the GP model. Fig. \ref{fig:rv_folded} compares the detection of TOI-6255 b using the GP and Floating Chunk Offset models. The GP model has far fewer parameters than the Floating Chunk Offset model and makes use of more data points. We adopted the GP model for all further analyses in this paper.

\section{Discussion}

\subsection{Core Compositions}
\label{sec:discussion_giant_impact}
With precise mass and radius measurements in hand, we can now investigate the possible compositions of TOI-6255 b. To put TOI-6255 b into context, we queried the NASA Exoplanet Archive \footnote{\url{https://exoplanetarchive.ipac.caltech.edu}} for the latest sample of terrestrial planets ($<1.8R_\oplus$) with well-measured masses and radii (both with $<20\%$ uncertainties). The sample is shown in Fig. \ref{fig:mass_radius}. We further identified a sub-sample of ultra-short-period planets (USP, $R_p<1.8R_\oplus$, $P_{\rm orb}<1$ day, blue symbols in Fig. \ref{fig:mass_radius}) to compare with TOI-6255 b. Previous studies \citep{USP,Lundkvist,Lopez2017,Kreidberg,Crossfield1252} have suggested that USP planets are so strongly irradiated by their host stars that any primordial H/He envelope should have been completely stripped away by intensive atmospheric erosion \citep{OwenWu,Ginzburg}. As such, one can directly probe the planet composition of these planets without worrying about a low-mean-molecular-weight atmosphere.  Indeed, \citet{Dai2019} have shown that the measured masses and radii of USPs are consistent with expectations for atmosphere-free rocky planets with an Earth-like composition of roughly 30\%Fe-70\% silicate rocks.

To fully characterize the composition of TOI-6255 b, we explore two models. The first model is a 2-layer interior model with an iron core and a silicate mantle \citep{Zeng2016}. The composition can be parameterized by a single number: the iron core mass fraction (CMF). We found that the CMF of TOI-6255 b is 0.45$\pm$0.32 using this model. Fig. \ref{fig:mass_radius} shows TOI-6255 b in comparison with other rocky exoplanets. We also employed the more sophisticated model  {\sc superearth} \citep{Valencia2006}.  {\sc superearth} further distinguishes between the true iron mass fraction (Fe-MF, Plotnykov \& Valencia, 2024) and the core mass fraction (CMF). This is to account for the fact that the silicate mantle may contain iron, while the iron core may contain other elements. Following the method in \citet{Plotnykov2020}, we included different degrees of core differentiation (expressed as the amount of iron in the mantle, $\mathrm{x_{Fe}}$), and different values for the Si alloy in the core ($\mathrm{x_{Si}}$). We use priors on these values based on the Terrestrial Planets $\mathrm{x_{Fe}}\sim U(0,20)$ \% by mol and $\mathrm{x_{Si}}\sim U(0,20)$ \% by mol. Using a Markov Chain Monte Carlo scheme, we found that TOI-6255 b has a CMF=0.41$\pm0.20$ and Fe-MF=0.38$\pm0.15$. These results are in agreement with the simpler model of \citet{Zeng2016}. A magma ocean may be present on TOI-6255 (see later discussion); \citet{Boley} showed that the presence of a magma ocean imposes negligible change on the inferred composition of a rocky planet from mass and radius measurement. On the other hand, the significant tidal distortion of TOI-6255 may affect the inference on its composition as will be explained in the next section.

\begin{figure}
\center 
\includegraphics[width = 1.1\columnwidth]{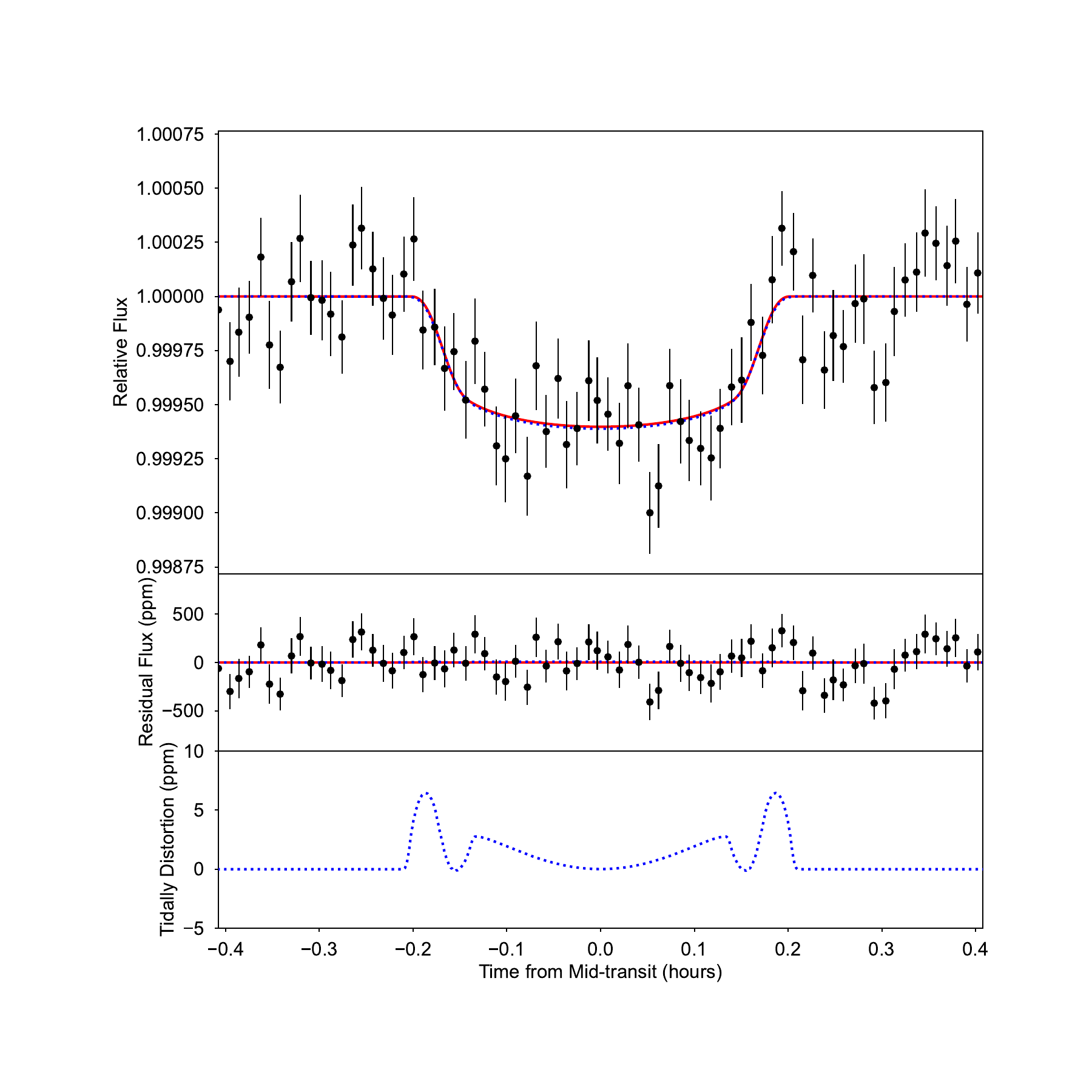}
\caption{ {\bf Top:} The transit of TOI-6255 b as observed by {\it TESS}. The black error bars are the phase folded and binned {\it TESS} light curve. The red curve is our best-fit transit model. The blue dotted line also includes the effect of tidal distortion. {\bf Middle:} The residual flux after removing the best-fit spherical planet model. {\bf Bottom:} The correction due to tidal distortion amounts to no more than 10 ppm for TOI-6255 b. This is much smaller than the {\it TESS} photometric precision and remains unconstrained.}
\label{fig:transit}
\end{figure}

\begin{figure*}
\center
\includegraphics[width = 1.9\columnwidth]{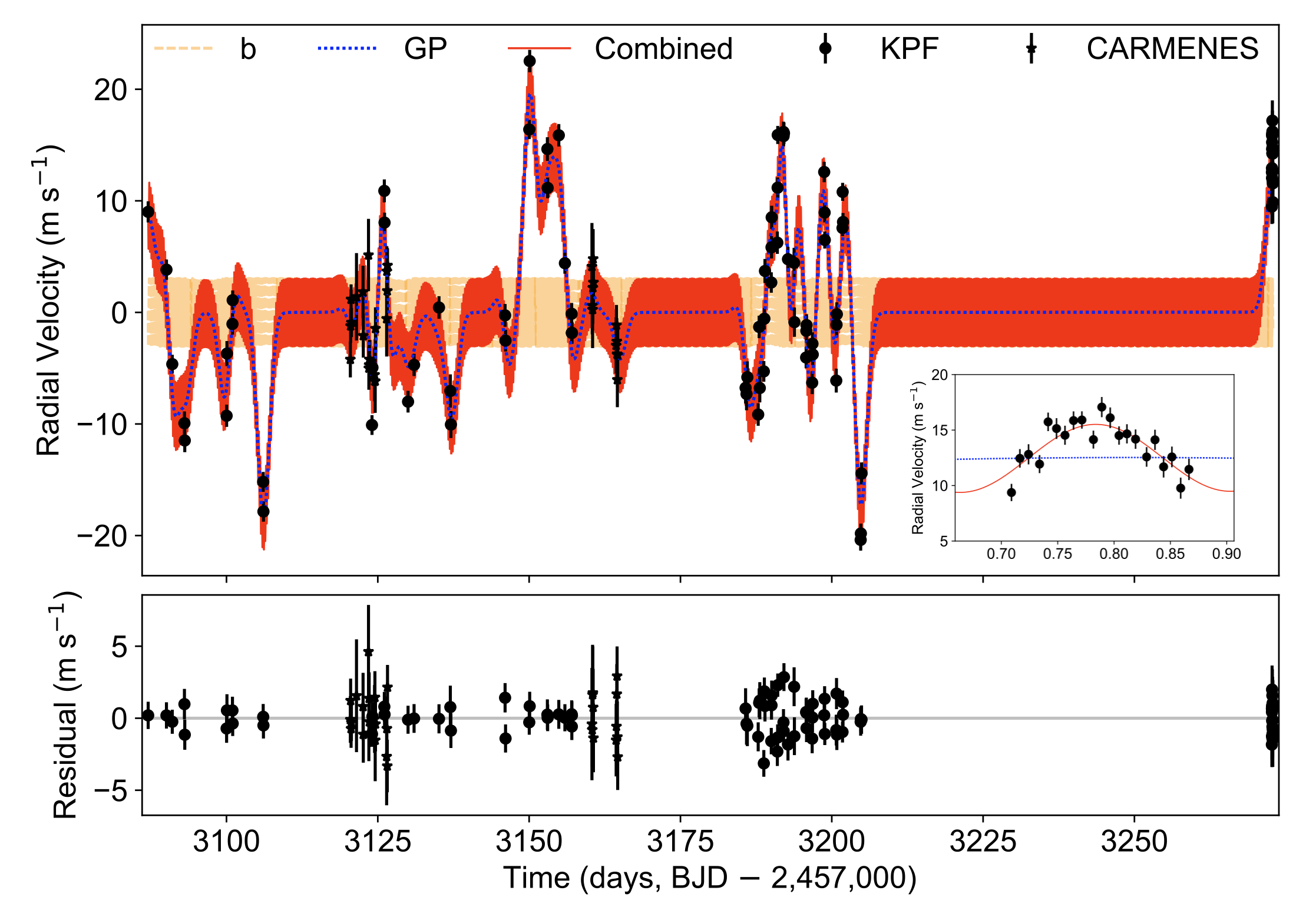}
\caption{Radial velocity (RV) variations of TOI-6255 b (orange dashed line) measured by KPF (circles) and CARMENES (stars). A Gaussian Process Model (blue dotted line) has been included to remove any stellar activity and instrumental variation on timescales longer than the orbital period of TOI-6255 b. On Nov 24, 2023, we took a set of 21 continuous exposures of TOI-6255 with KPF (inset). The resultant RVs followed clearly reveal the 5.7-hour orbit of TOI-6255 b. }
\label{fig:rv}
\end{figure*}

\begin{figure}
\center
\includegraphics[width = 1.\columnwidth]{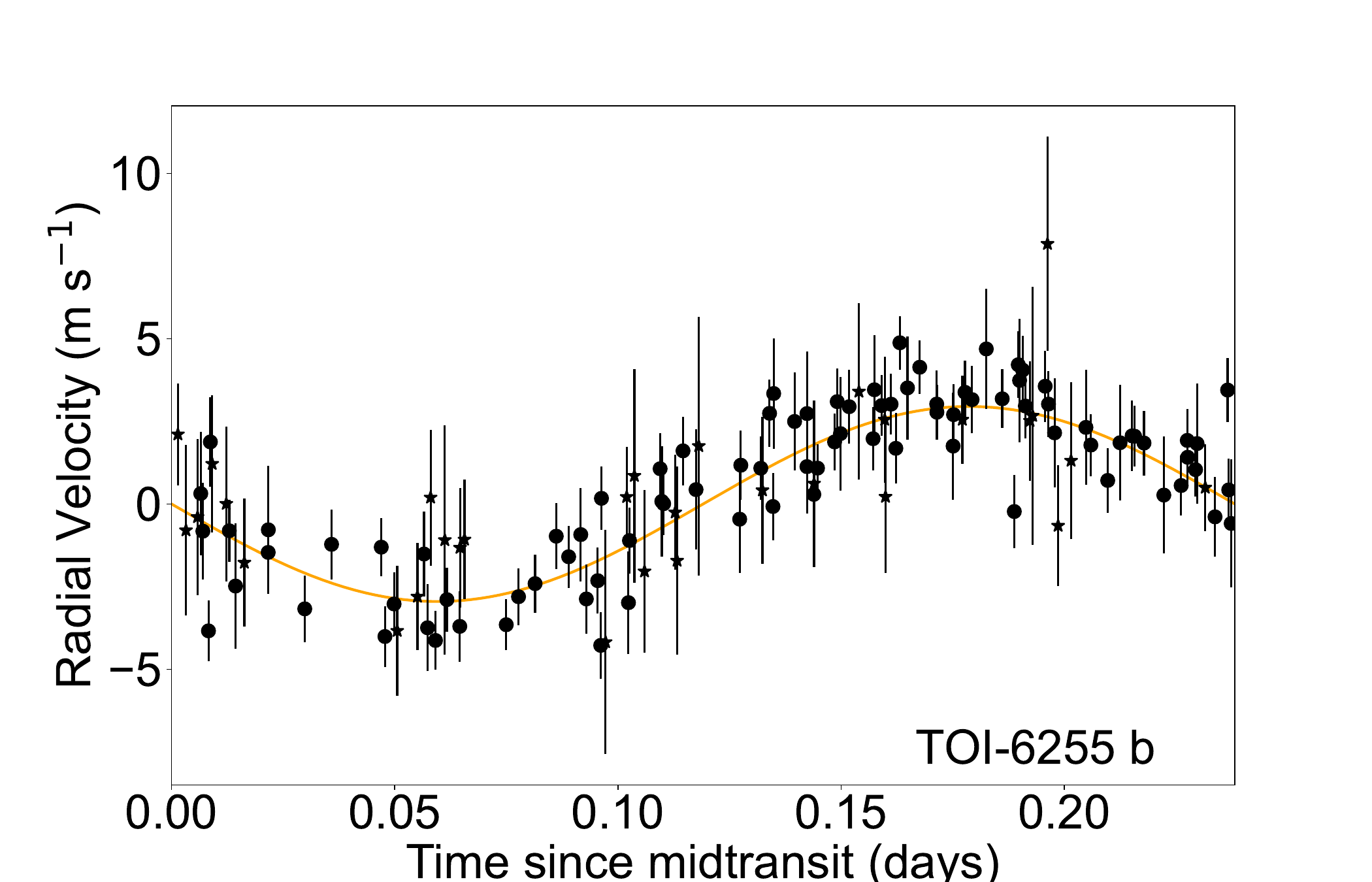}
\includegraphics[width = 1.\columnwidth]{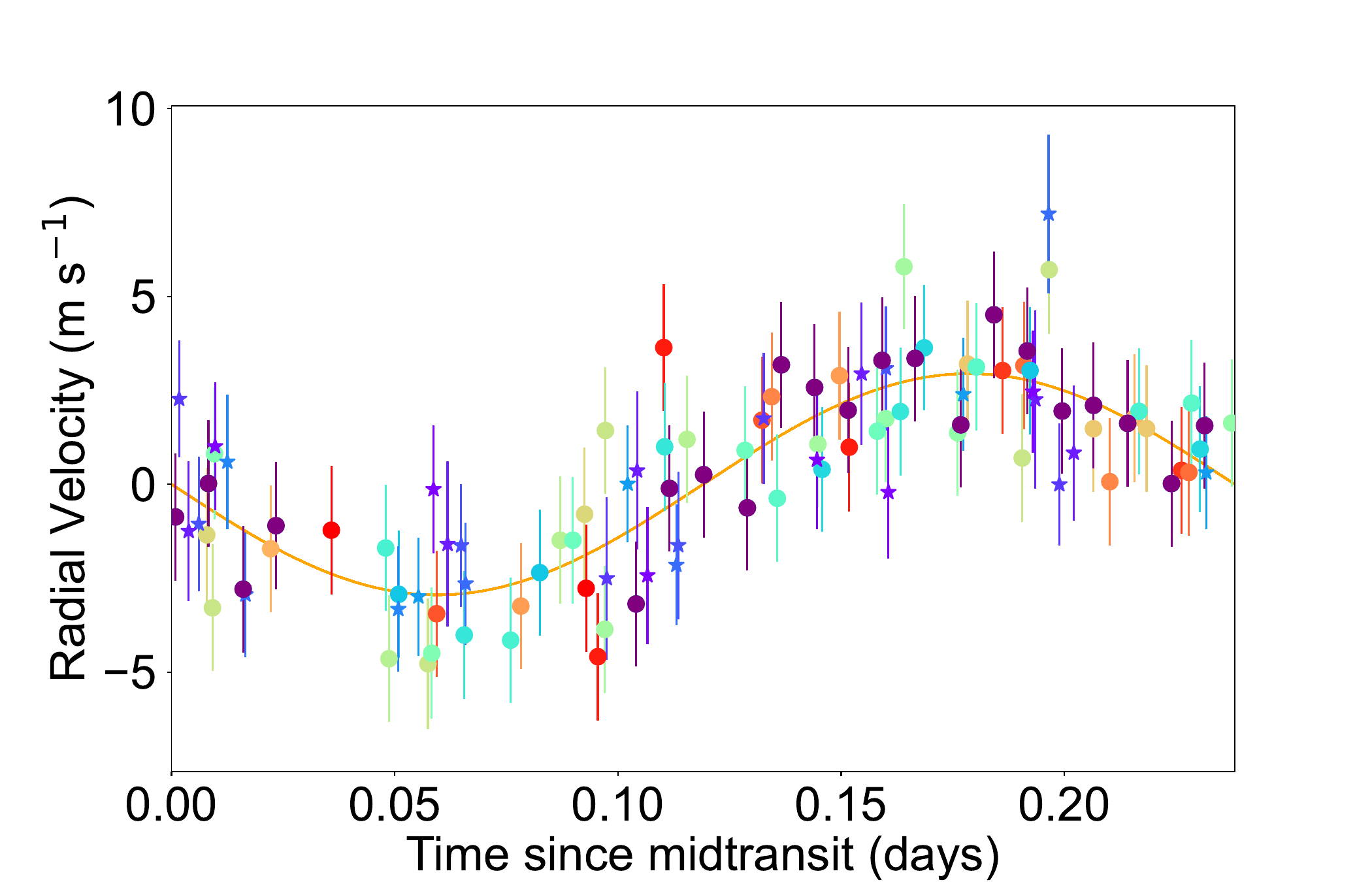}
\caption{ The radial velocity variation of TOI-6255 b plotted as a function of the orbital phase. {\bf Top}: Results of our Gaussian Process model where long-term stellar activities and instrumental effects have been removed before plotting. {\bf Bottom}: An alternative model (Floating Chunk Offset, Section \ref{sec:fco}) where observations taken on a single calendar night were given an independent offset (plotted with a different color). These two models give consistent results on the radial velocity semi-amplitude of TOI-6255 b ($K=2.98\pm0.28$m~s$^{-1}$ v.s.
$K=2.93\pm0.38$m~s$^{-1}$).}
\label{fig:rv_folded}
\end{figure}

\begin{figure*}
\center
\includegraphics[width = 1.8\columnwidth]{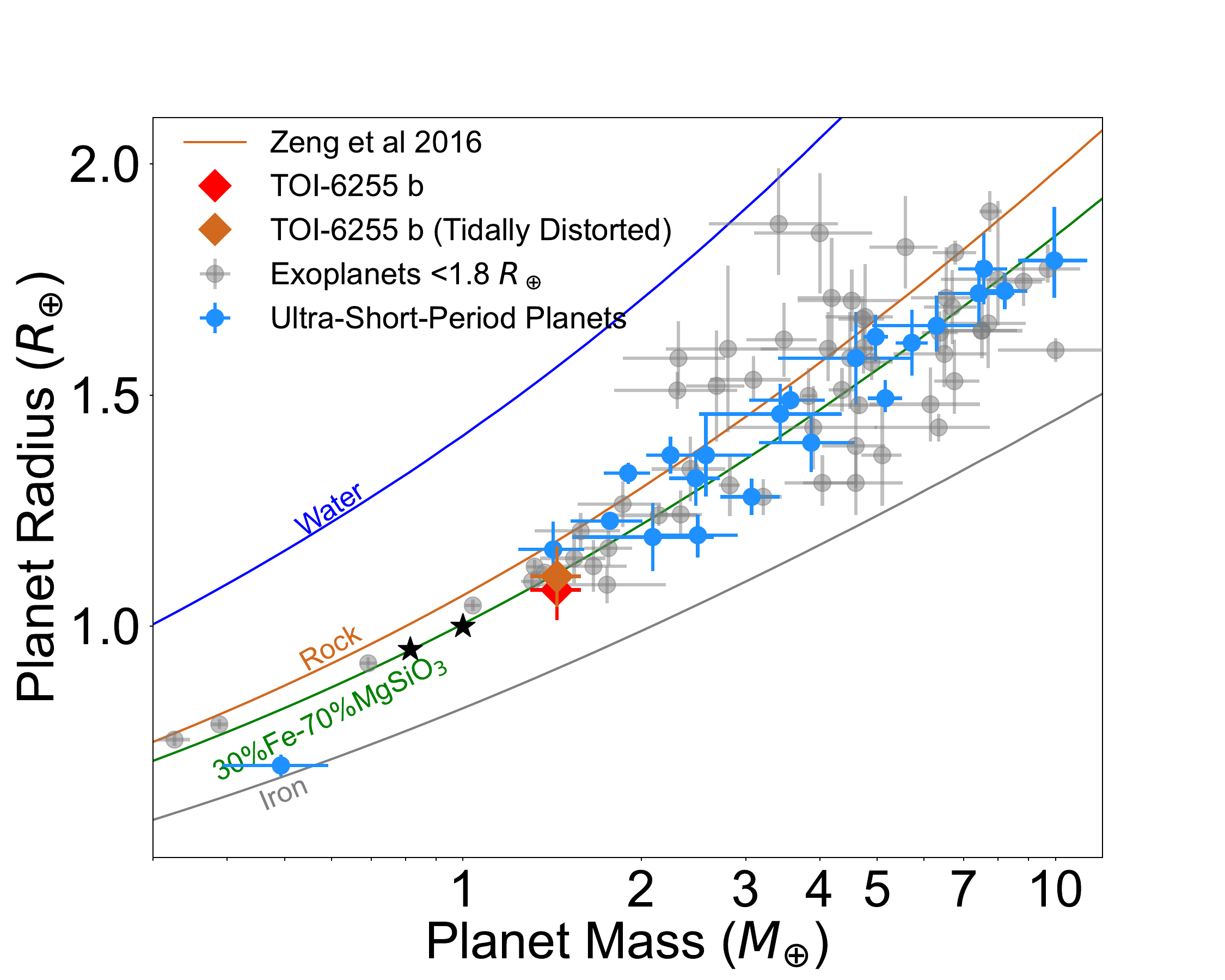}
\caption{Precise mass and radius measurements of exoplanets $<1.8R_\oplus$ (from NASA Exoplanet Archive) and theoretical mass-radius relationships \citep{Zeng2019}. The ultra-short-period planets ($P_{\rm orb}<1$ day, $R_p<1.8R_\oplus$ blue points) are so strongly irradiated that they should be bare rocky bodies devoid of H/He envelopes. The existing sample of ultra-short-period planets are dominated by super-Earths ($>2M_\oplus$). As an ensemble, they cluster around an Earth-like 30\%Fe-70\%MgSiO$_3$ composition \citep{Dai2019}.  TOI-6255 b is in the domain of Earth-sized planets. TOI-6255 b has a CMF of $45\pm32$\% (red diamond).  If accounting for the tidal distortion with a Love number $h_2=1$, the inferred CMF drops to CMF of $31\pm30$\% because during transit the shorter two axes of the ellipsoidal planet are visible (brown diamond). See Section \ref{sec:discussion_distortion} for details.}
\label{fig:mass_radius}
\end{figure*}

\begin{figure*}
\center
\includegraphics[width =1.5\columnwidth]{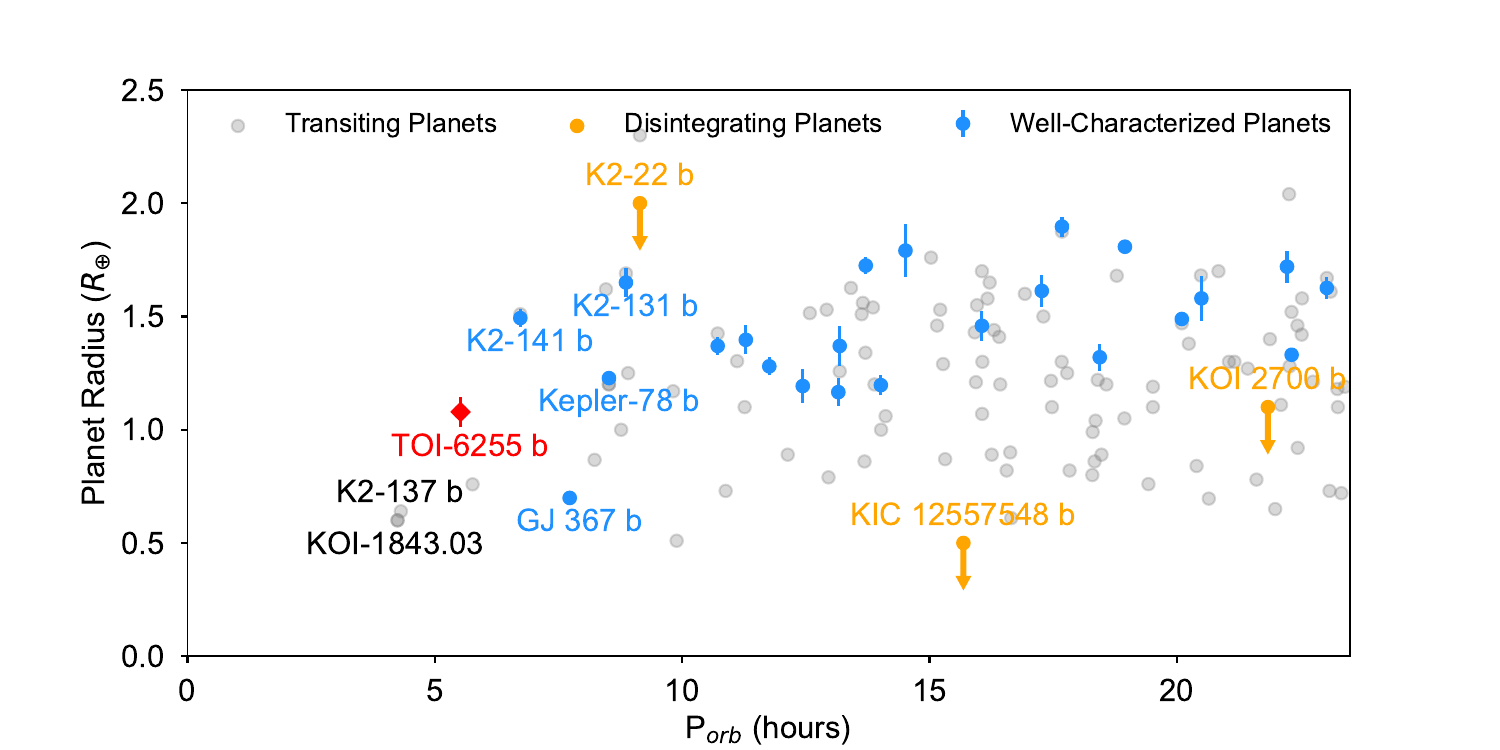}
\includegraphics[width = 1.5\columnwidth]{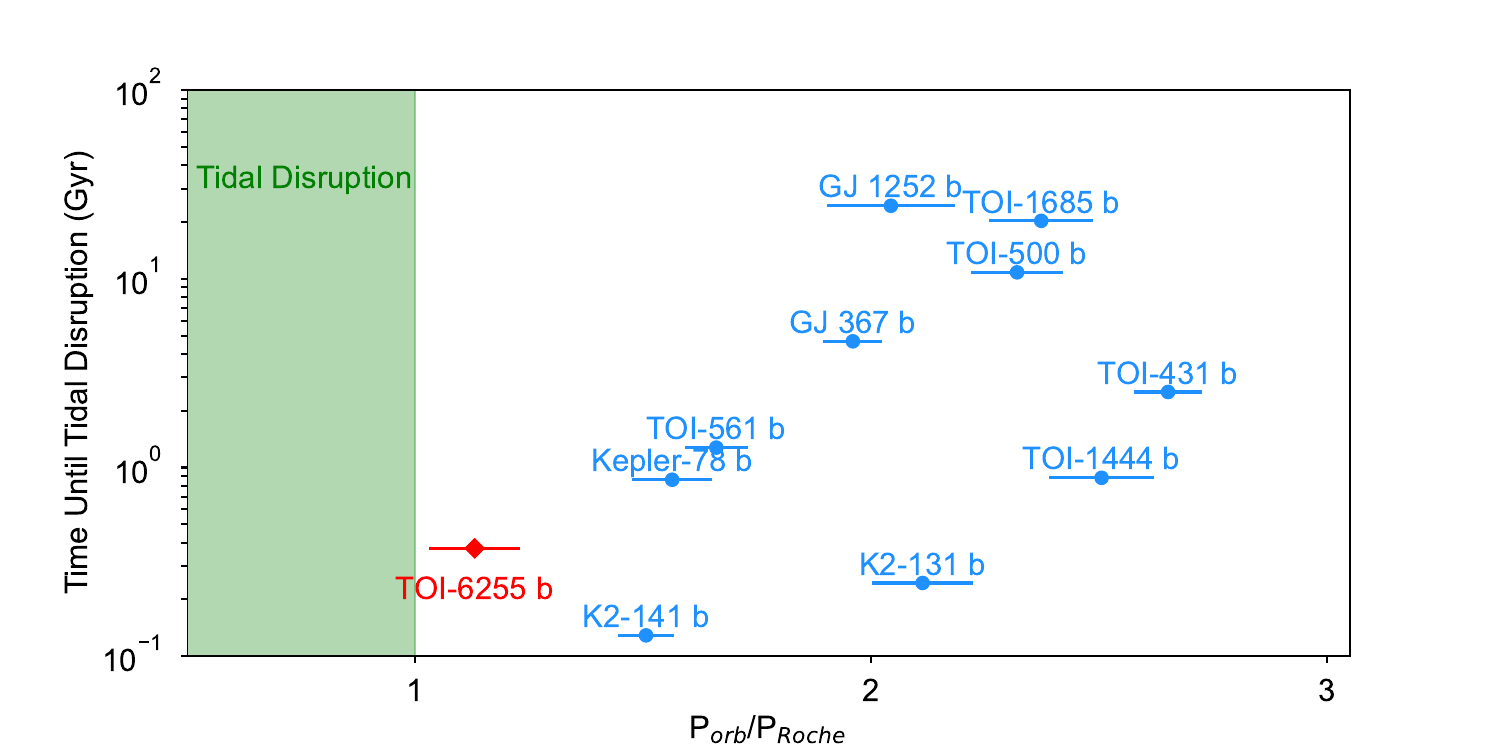}
\includegraphics[width = 1.5\columnwidth]{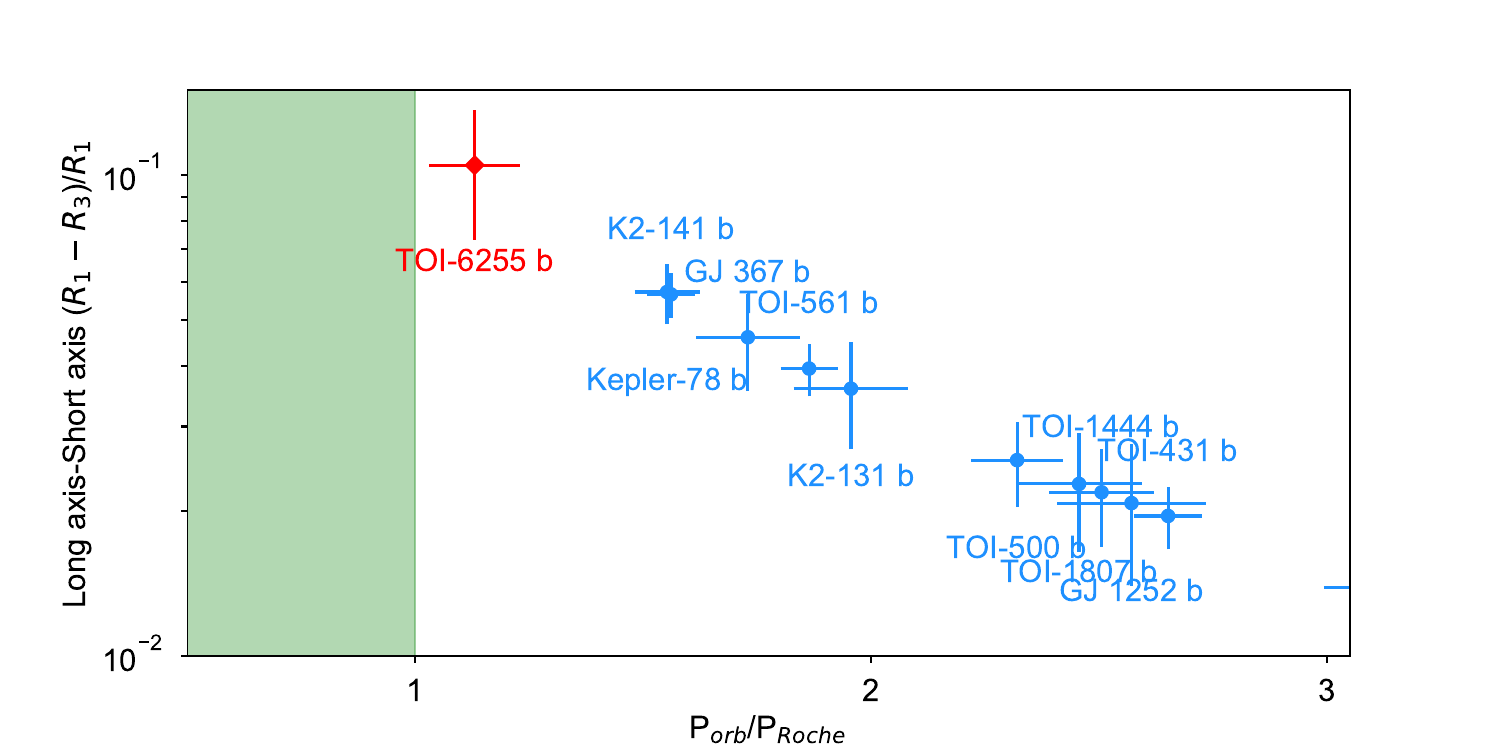}
\caption{{\bf Top:} With an orbital period of 5.7 hours and a semi-major axis ($a/R_\star=3.1$), TOI-6255 b is close to the tidal disruption limit. K2-137 b \citep{Smith} and KOI-1843.03 have shorter orbital periods. However, the masses of those planets are unknown. {\bf Middle:} Using the equation for Roche limit $P_{\rm Roche} \approx 12.6 ~{\rm hours} \times(\rho_p/1$ g~cm$^{-3}$)$^{-1/2}$ from \citet{Rappaport}, one can compute how far a planet is away from tidal disruption $P_{\rm orb}/P_{\rm Roche}$. TOI-6255 b stands out as being on the verge of tidal disruption with a $P_{\rm orb}/P_{\rm Roche}=1.13\pm0.10$. The y-axis is how fast the planet would reach the Roche Limit due to equilibrium stellar tides (Eqn \ref{eqn:tidal_decay}, notice the strong dependence on orbital distance). TOI-6255 b may experience the fate of tidal disruption in the next 400 Myr if the tidal quality factor of the star $Q_\star^\prime=10^7$. Error bar is at least an order of magnitude due to uncertainty on $Q_\star^\prime$. {\bf Bottom:} Tidal distortion (the fractional difference between the long axis and the short axis of the ellipsoid) as predicted by Tidal Love theory (Eqn. \ref{eqn:tidal_distortion}) with a Love number $h_2=1$. TOI-6255 b is likely the most tidally distorted terrestrial planet discovered so far.}

\label{fig:tidal}
\end{figure*}

\subsection{Tidal Distortion}\label{sec:discussion_distortion}

Given the extremely short orbital period of TOI-6255 b, it may be significantly tidally deformed. To demonstrate that TOI-6255 b is under extreme tidal stress, we first compared its orbital period to the Roche limit of an incompressible fluid:
\begin{equation}
P_\mathrm{Roche} \approx 12.6~\mathrm{h} \left(\frac{\rho_p}{1~\mathrm{g}~\mathrm{cm}^{-3}}\right)^{-1/2},
\end{equation}
as in \citet{Rappaport}.  $P_\mathrm{Roche}$ is the orbital period of the Roche Limit where tidal forces due to the star are stronger than the gravity of the planet i.e. the planet starts to disintegrate if material strength is negligible. The orbital period of TOI-6255 b is dangerously close to the Roche limit with $P_\mathrm{orb}/P_\mathrm{Roche}$ = 1.13$\pm0.10$. Fig. \ref{fig:tidal} shows that although dozens of ultra-short period planets have been discovered, TOI-6255 b stands out as the planet that is closest to tidal disruption.

 One can estimate the extent of tidal distortion of the planet using the Love number $h_2$ \citep{Love1944}. For a homogeneous solid planet:

\begin{equation}\label{eqn:tidal_distortion}
    \delta R_{p}=h_2\zeta
\end{equation}

\begin{equation}
    h_2 = \frac{5/2}{1+\tilde{\mu}}
\end{equation}

\begin{equation}
    \tilde{\mu} = \frac{19\mu}{2\rho g R_{p}}
\end{equation}

\begin{equation}
    \zeta= \frac{M_{\star}}{M_{p}}\left(\frac{R_{p}}{a}\right)^3R_{p}
\end{equation}

\noindent where $R_{p}$ and $\delta R_{p}$ are the nominal radius and tidal distortion of the planet $M_{p}$ and $M_{\star}$ are the masses of the planet and the host star. $\mu$ the tensile strength of the planet. $\rho$ is the mean density of the planet.  $g$ is the surface gravity of the planet.  $a$ is the semi-major axis.  These quantities have been directly measured with the exceptions of $h_2$ and $\mu$. For a centrally concentrated planet, $h_2$ is further reduced from $\frac{5/2}{1+\tilde{\mu}}$. \citet{Kellermann} showed that increasing the iron core mass fraction of an Earth-sized planet from 0 to Earth-like (30\%) leads to roughly a factor 2 reduction in the tidal Love number. As for the tensile strength $\mu$, \citet{Murray} recommend a mean tensile strength of 50 GPa for rocky bodies within the solar system; this would further reduce $h_2$ by a factor of $(1+\tilde{\mu})\approx 1.9$. However, for TOI-6255 b the high equilibrium temperature ($\sim1300$K) likely leads to at least partial melting and hence significant weakening of the tensile strength of the surface rocks \citep{Takahashi}. \citet{Saiang} showed that from room temperature to 1000 K, the tensile strength of common rocks can easily drop by an order of magnitude. Moreover, since the planet is likely in synchronous rotation and on a circular orbit (see Section \ref{sec:decay} for tidal despin and circularization timescale), the tidal bulge on the planet is static in the corotating frame centered on the planet. Solid body thermal creep (again equilibrium temperature is high enough) might have adjusted the planet to a hydrostatic equilibrium by slowly releasing any internal stress. Thus, $\tilde{\mu}$ can be negligible on TOI-6255 b. Empirically, estimates of Earth's Love number $h_2$ range between 0.6 and 0.9 \citep[e.g.][]{Lambeck_1980,Krasna}. For TOI-6255 b, $h_2$ is likely higher than Earth given the weakened material strength and larger planetary size.  We adopt a fiducial $h_2=1$ for further analysis. 

\begin{figure*}
    \centering
    \includegraphics[width=0.48\linewidth]{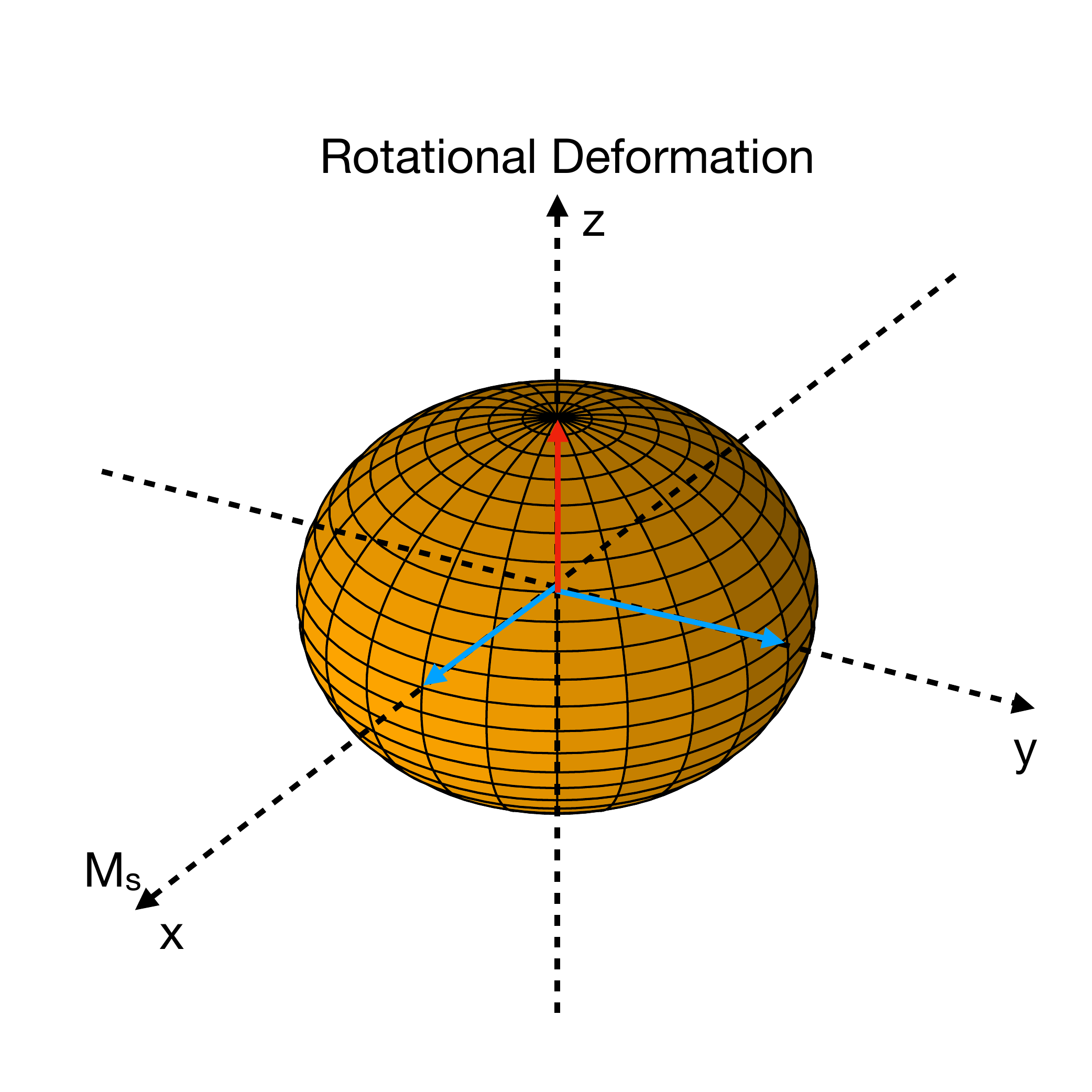}
    \includegraphics[width=0.48\linewidth]{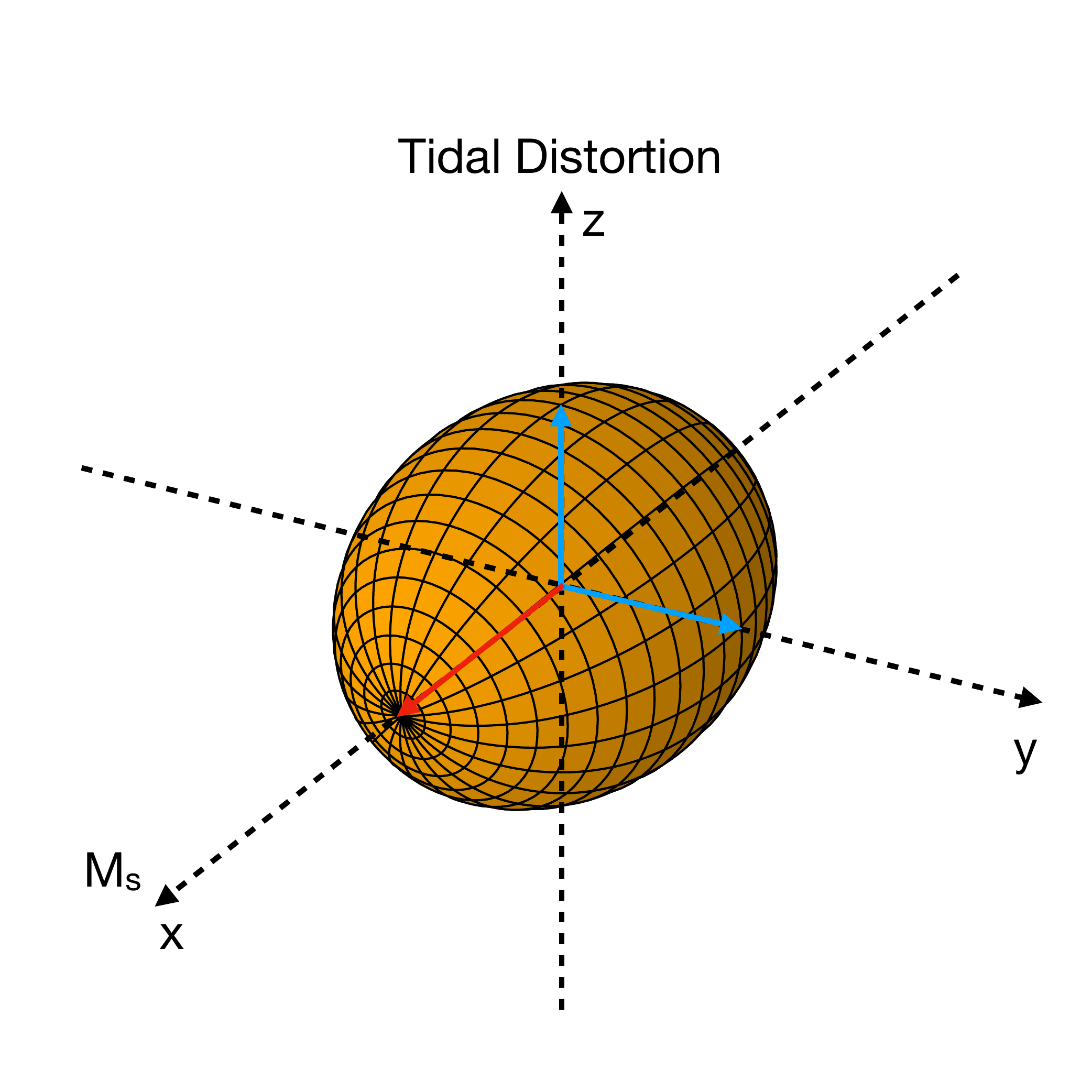}
    \includegraphics[width=0.48\linewidth]{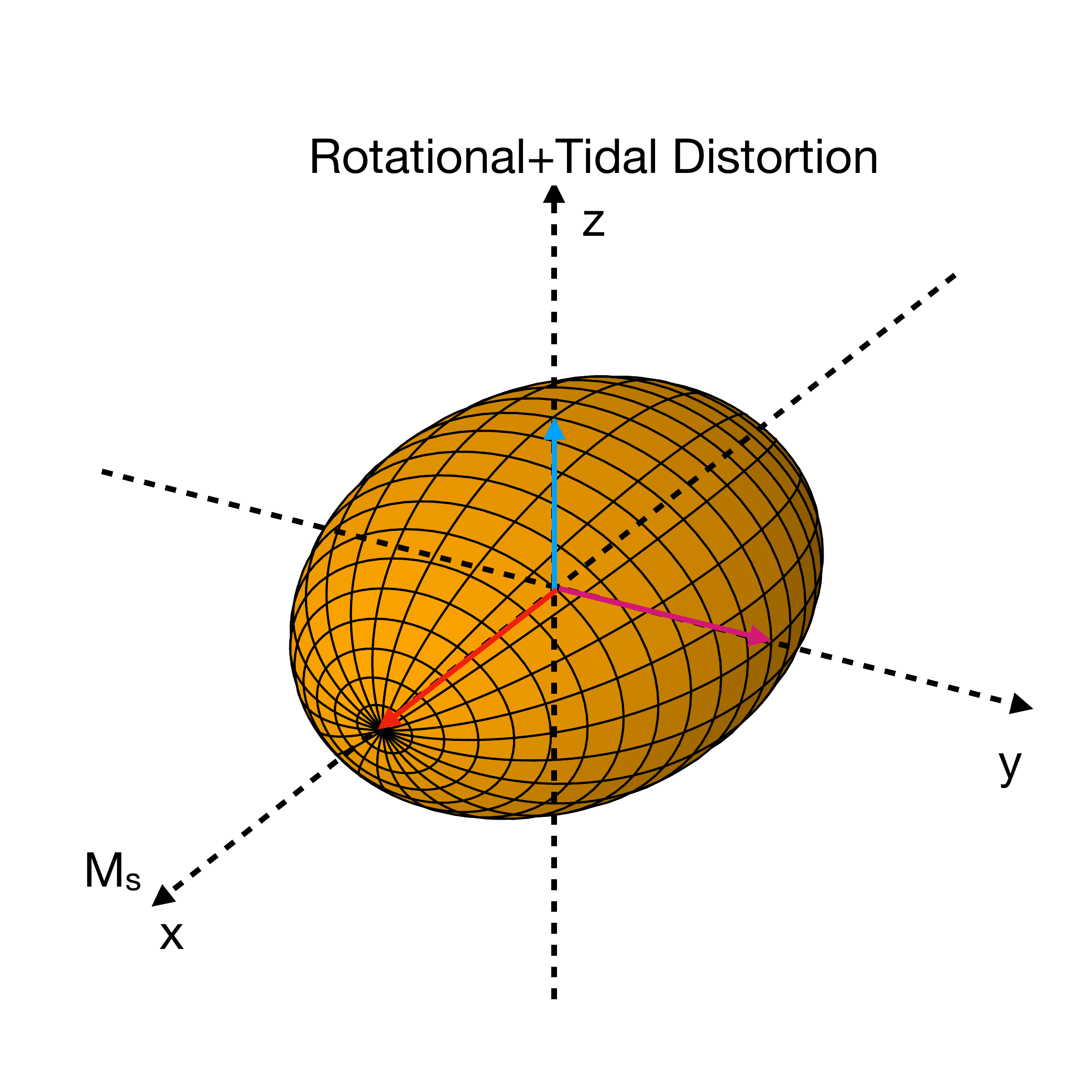}
    \caption{A schematic showing the rotational deformation (top left), tidal distortion (top right), and the combined effect for a synchronously rotating planet (bottom). The host star is located along the x-axis. The orbital motion is against the y-axis direction. The amount of distortion has been exaggerated for visualization. The red, magenta, and blue arrows indicate the long, intermediate, and short semi-axis of the ellipsoid ([$R_1$,$R_2$,$R_3$]). Rotational deformation produces an oblate planet with a bulge at the equator (see Eqn \ref{eqn:rotational_deformation}). Tidal distortion is along the axis joining the planet and the host star (see Eqn \ref{eqn:tidal_distortion}
    ).}
    \label{fig:tidal_distortion}. 
\end{figure*}

The 5.7-hour rotation period of TOI-6255 also leads to substantial rotational deformation:

\begin{equation}\label{eqn:rotational_deformation}
    q = \frac{\Omega^2 R_{p}^3}{G M_{p}}
\end{equation}
where $\Omega$ is rotational angular velocity of the planet, $G$ is the gravitational constant. $q$ quantifies surface gravity against the centrifugal force at the surface of the planet.  $q$ amounts to about 4\% on TOI-6255 b. For a synchronously rotating planet, it can be shown that the effective potential of the rotation is $-1/3$ of the tidal potential \citep{Murray}. The tidal bulge is along the line connecting the host star and the planet (assuming a negligible tidal lag), while the rotational deformation bulges out the equator of the planet. The addition of these two effects results in an ellipsoidal shape (with semi-major axes [$R_1$,$R_2$,$R_3$]) as illustrated in Fig. \ref{fig:tidal_distortion}. The long axis $R_1$ points towards the host star. The intermediate axis $R_2$ points along the direction of orbital motion. The short axis $R_3$ is the polar axis parallel to the rotation axis of the planet. The three semi-axes of the ellipsoids are given by \citep[see e.g.][]{Correia2014}:

\begin{equation}
   R_1=R_{\rm vol}(1+\frac{7}{6}\delta R_{p})
\end{equation}
\begin{equation}
   R_2=R_{\rm vol}(1-\frac{1}{3}\delta R_{p})
   \end{equation}
\begin{equation}
   R_3=R_{\rm vol}(1-\frac{5}{6}\delta R_{p})
\end{equation}

\noindent where $R_{\rm vol} \equiv (R_1R_2R_3)^{1/3}$ is the volumetric radius of the planet. A tidally distorted planet should also produce a distinct transit signal compared to a spherical model \citep{Leconte}. In particular, if the planet is in synchronous rotation, the long axis of the planet always points towards the star. Hence, the shadow of the planet projected on the star changes as a function of the orbital phase and orbital inclination: 

\begin{equation}\label{eqn:projected_area}
   A(\phi) = \pi \sqrt{R_1^2R_2^2\cos^2{i}+R_3^2\sin^2{i}(R_1^2\sin^2{\phi}+R_2^2\cos^2{\phi})}
\end{equation}

for a transiting planet during transit, the orbital inclination $i$ is close to 90$^\circ$ and the  orbital phase $\phi$ is close to 0.  $R_1$ is mostly hidden from the observer. Therefore, the transit radius of the planet is roughly given by $R_{\rm tran} \equiv (R_2R_3)^{1/2}$. As such, the volumetric radius of the planet is larger than the transit radius of the planet: $R_{\rm vol} = R_{\rm tran}(1+\frac{7}{12}\delta R_{p})$.

We now revisit the composition of TOI-6255 b if we account for both tidal and rotational deformation. We can obtain the volumetric radius of the planet from its transit radius and the expected distortion. We then used the volumetric radius to infer the composition, primarily the iron CMF, of TOI-6255 b. Assuming $h_2$ = 1, this resulted in roughly a 3\% increase in the transit radius. The inferred CMF of the planet dropped from 45$\pm32$\% to 31$\pm30$\%. Such a correction does not change the qualitative result presented in Section \ref{sec:discussion_giant_impact}. 

In principle, the ellipsoidal shape of the planet can be directly measured in the transit light curve and phase curve variation. An ellipsoidal planet produces two major effects in the transit light curve: 1) it gives rise to an extension of the ingress/egress timescale, the tidal bulge will occult the star before the nominal ingress of spherical planet, and 2) the tidal bulge is hidden from observer's view at mid-transit, but away from mid-transit it slowly rotates into view and causes extra absorption compared to a spherical planet. We looked for both effects in the {\it TESS} light curve by explicitly computing the transit light curve of an ellipsoidal planet (lower panel of Fig. \ref{fig:transit}). The light curve correction due to an ellipsoidal planet is $<10$ppm. The binned uncertainty of the {\it TESS} light curve on a 2-min timescale (comparable to the ingress/egress timescale) is about 120ppm. We could not robustly detect the signature of tidal distortion in the {\it TESS} light curve.  \citet{Hellard} performed a detail analysis on detectability of tidal distortion in {\it TESS} data, they found that such a detection is challenging even for giant planets.

The calculation of the tidal and rotational deformation presented so far assumes small deformation and linear responses. In other words, the deformations are small compared to the size of the planet, and the different effects can be simply summed linearly. However, the tidal distortion on TOI-6255 ($\sim$10\%) is probably no longer linear i.e. beyond the yield point of most solids \citep[e.g. steel at $\sim0.1\%$][]{ross1999mechanics}. In comparison, the lunar tide on Earth is of order $10^{-7}$. 

Previously, \citet{Price} expanded on the self-consistent field method of \citet{Hachisu1986a,Hachisu1986b} to compute the tidal distortion of small, rocky planets in three dimensions, taking into account gravitational forces from the star and planet and the centrifugal force from the planet's rapid orbital motion. The model allows the planet to have an arbitrary number of layers. The \citet{Hachisu1986a,Hachisu1986b} method is a relaxation procedure for solving the integro-differential equation governing the planet's equilibrium shape. In practice, the method rapidly converges to the equilibrium solution, within a specified tolerance, in just a few iterations.

We applied the model of \citet{Price} on TOI-6255 b. A potential limitation of the model by \citet{Price}  is that parameters of scientific interest such as orbital period, planet mass, and core mass fraction are outputs and cannot be specified upfront. For our purposes, we generated a library of models spanning a wide range of core-mantle boundary pressures, core pressures, aspect ratios, and planet-star distances. Within this model library, we drew and interpolated models that are broadly consistent with the measured stellar mass, planet mass, orbital period, and transit radius of TOI-6255 b. The results suggest that TOI-6255 b may indeed be more significantly distorted than what the Love theory predicts. The long axis may be 15\% longer than the short axis. The model by \citet{Price} should be valid in the non-linear regime of tidal distortion; however, it does not include material strength yet. In Love's linear theory of tidal distortion, material strength is included (see Eqn. \ref{eqn:tidal_distortion}). It is not clear which model offers a better description of the tidal distortion of TOI-6255 b; nonetheless, the two models agree that TOI-6255 b is one of the most tidally distorted terrestrial planets discovered so far. We defer a detailed modeling of the tidal distortion when it can be empirically constrained.

\begin{figure}
\center
\includegraphics[width = 1.\columnwidth]{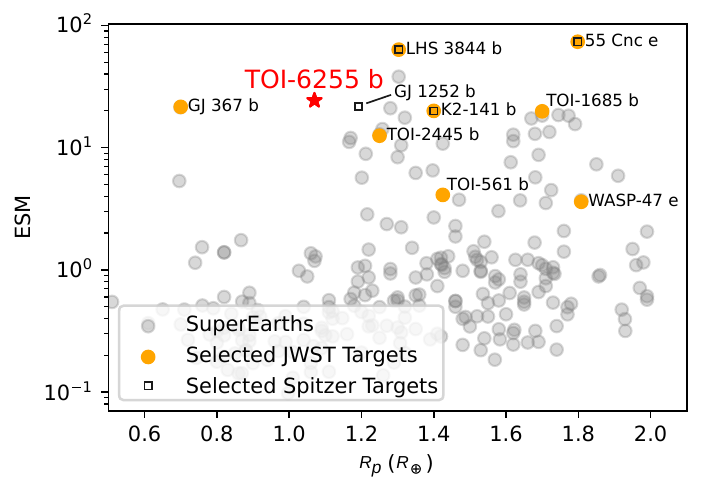}
\caption{The Emission Spectroscopy Metric \citep[ESM, ][]{Kempton} v.s. planet radius, for
planets smaller than $2\,R_\oplus$ (super-Earths). TOI-6255 b has an ESM of 24, and hence is more favorable than many planets for which JWST observations have already been scheduled (orange). It is also the most Earth-like in size among the currently confirmed exoplanet sample.}
\label{fig:esm}
\end{figure}

\subsection{Tidal Decay and Dissipation} 
\label{sec:decay}

We have seen that TOI-6255 b may be tidally distorted due to its proximity to the Roche limit. Is it in any danger of being tidally disrupted in the near future? Using the constant-lag-angle model of equilibrium tides \citep{Goldreich1966,Winn2018}, the tidal decay timescale due to tides raised on the slowly rotating host star is given by 

\begin{equation}\label{eqn:tidal_decay}
\frac{P}{\Dot{P}}\approx 30\,{\rm Gyr} \left(\frac{Q_\star^\prime}{10^6}\right)\left(\frac{M_\star/M_p}{M_\odot/M_\oplus}\right)\left(\frac{\rho_\star}{\rho_\odot}\right)^{5/3}\left(\frac{P_{\rm orb}}{1\,{\rm day}}\right)^{13/3}
\end{equation}
where $Q_\star^\prime$ is the tidal quality factor for the host star. $M_p$ and $M_\star$ are the masses of the planet and the host star. $\rho_\star$ is the mean stellar density. Using a nominal $Q_\star^\prime=10^7$ for the host star, TOI-6255 b has a tidal decay timescale of $\tau_P\equiv P/\Dot{P }\approx$3.7 Gyr. However, the orbital period only needs to change by $\sim$ 13\% to reach the Roche Limit, which takes about $\tau_{\rm Roche}\approx$ 400 Myr if integrating Eqn. \ref{eqn:tidal_decay} (note the difference between $\tau_P$ and  $\tau_{\rm Roche}$). The major source of uncertainty of this calculation comes from the tidal quality factor of the host star $Q_\star^\prime$ which may vary by more than an order of magnitude and may be frequency-dependent as suggested by recent theoretical and empirical results \citep[e.g.][]{ogilvie2014, Weinberg,Bailey,Penev}. $Q_\star^\prime=10^7$ represents a relatively slow stellar dissipation rate \citep{Penev}.  TOI-6255 is likely a mature M-star with an age of several Gyr as evidenced by the slow rotation. Tidal perturbation by a USP around a slowly spinning host stars is outside the inertial 
range rendering the suppression of dynamical tide \citep{ogilvie2007}.

A priori, it is unlikely that a planet would be discovered right before its destruction. As shown in Section \ref{sec:stellar_para},
400 Myr represents about 5-10\% of the system's age. More than a hundred USPs have been confirmed in the literature. In this light, it seems possible that the sample contains a planet whose tidal decay timescale is 5-10\% of the system's age. Fig. \ref{fig:tidal} shows the time required for TOI-6255 b and other shortest-period planets to reach the tidal disruption limit $\tau_{\rm Roche}$. Among the USPs ($P_{\rm orb}<1$ day), the timescales can be shorter than the age of the universe or the age of the host star. Once the planet is disrupted, the resultant debris is likely to fall onto the host star. Depending on how permeable the radiative-convective boundary is, the refractory elements may linger in the surface convective layer. The photospheric composition of a star would thus be enhanced in refractory elements found in rocky planets. By comparing the abundance of refractory elements on the photospheres of comoving, co-natal binaries, it has been suggested that the star that is preferentially enriched in refractory elements experienced engulfment of rocky planets  \citep{Ram,Oh,Nagar,Galarza,Spina,Behmard,Liu2024}. TOI-6255 b may be a progenitor of such a catastrophic planet engulfment event.

If TOI-6255 b still has a non-zero orbital eccentricity, tidal dissipation may take place in the planet. If so, the orbital decay rate can be sped up by including an additional term due to the tidal dissipation in the planet:

\begin{equation}
\begin{aligned}
& \ \frac{P}{\Dot{P}}\approx 2 {\rm Gyr} \left(\frac{Q_p^\prime}{10^3}\right)\left(\frac{M_p}{M_\oplus}\right)\left(\frac{M_\star}{M_\odot}\right)^{2/3}\\
 &\quad \left(\frac{R_p}{R_\oplus}\right)^{-5}\left(\frac{P_{\rm orb}}{1 {\rm day}}\right)^{13/3}\left(\frac{e}{0.01}\right)^{-2}
\end{aligned}
\end{equation}

 Using a nominal planetary tidal quality factor $Q_p^\prime=1000$ similar to Mars \citep{Murray} and an $e$ of $0.001$ level \citep[c.f. Io's eccentricity is about 0.004][]{Peale_io}, the tidal decay timescale for TOI-6255 b is about $\tau_P \approx200$ Myr. The tidal dissipation amounts to $\sim7\%$ of the insolation the planet receives. Again the major source of uncertainty is in $Q_p^\prime$ whose uncertainty can be more than an order of magnitude. \citet{Peale_io} pointed out that as the planet's core gets molten, the solid mantle gets increasingly thinner. The mantle experiences stronger deformation and thus stronger dissipation. On the other hand, if the planet's interior is completely molten (by tidal heating or by star-planet magnetic interaction Section \ref{sec:star-planet}), the fluid-like interior is much less dissipative. $Q_p^\prime$ can be orders of magnitude higher, similar to the giant planets.  
 
 Equally important, can TOI-6255 b maintain a non-zero eccentricity? The tidal circularization time is given by: 
\begin{equation}
\begin{aligned}
& \ \tau_e\equiv\frac{e}{\Dot{e}}\approx 0.7\,{\rm Myr} \left(\frac{Q_p^\prime}{10^3}\right)\left(\frac{M_p/M_\star}{M_\oplus/M_\odot}\right)\\
&\left(\frac{R_\star/R_p}{R_\odot/R_\oplus}\right)^5\left(\frac{\rho_\star}{\rho_\odot}\right)^{5/3}\left(\frac{P_{\rm orb}}{1\,{\rm day}}\right)^{13/3}
\end{aligned}
\end{equation}
 For TOI-6255 b, this is only $\tau_e\approx$ 400\,yr if $Q_p^\prime$ is 10$^3$. We briefly note that the tidal despin timescale for TOI-6255 b to become tidally locked is even shorter by a factor $(R_p/a)^2$, and can be shorter than one year for TOI-6255 b at its current orbital period. Such a short timescale justifies our earlier assumption that the planet is tidally locked with a constant day side and night side.

Given the short tidal circularization timescale, the planet should have a circular orbit unless the other planets constantly pump up the eccentricity through resonant or secular interaction \cite[e.g.][]{Schlaufman_usp,Petrovich,Pu2019}. Previous works have shown that USP planets like TOI-6255 b are almost always found in multi-planet systems \citep{Winn2018,Dai_1444}. \citet{Hansen} proposed a scenario for 55 Cnc e in which the planet maintains a non-zero eccentricity due to the perturbation of other planetary companions. We are currently unable to confirm any additional planets around TOI-6255 that could pump up the eccentricity of TOI-6255 b. Without an orbital eccentricity, most of the tidal dissipation happens within the host star as described by Eqn \ref{eqn:tidal_decay}.

\subsection{Star-Planet Magnetic Interaction}\label{sec:star-planet}

It has long been suspected that ultra-short-period planets like TOI-6255 b may be within the stellar Alfv\'{e}nic sphere and experience a strong star-planet magnetic interaction \citep[e.g.][]{Cuntz_2000}. The interaction is much analogous to our Io-Jupiter system \citep{Bigg}. The transfer of ions from the planet back to the star, through a structure often referred to as the Alfv\'{e}n wings, may lead to radio emission from the system \citep{Zarka1998}. The associated Ohmic dissipation may also give rise to orbital decay and melting of the interior of the planet \citep[in addition to the tidal effect, ][]{Laine, Wei2024}. TOI-6255 b is a top candidate for detecting such star-planet magnetic interactions.  

The deca-meter radio emission from Jupiter is likely due to the electron cyclotron maser (ECM) instability \citep{Melrose1982,Zarka1998}. Star-planet magnetic interactions are capable of generating auroral radio emission from both the star and the planet. This non-thermal emission would have a characteristic gyro-frequency of $\nu_G = 2.8 \, B$ MHz, where $B$ is the local magnetic field (in Gauss). The emission would be coherent, broad-band($\Delta\,\nu \sim \nu_G/2$), and highly circularly polarized (probably reaching close to 100\% in some cases).  Moreover, the stellar radio emission may be modulated by the orbital phase of the planet just like Jupiter's decameter emission.

We note that any potential emission coming from the planet would most likely be undetectable, as the magnetic field strengths of rocky planets in the solar system do not exceed a few Gauss. The resulting ECM frequency is below Earth's ionosphere cutoff. On the other hand, the stellar magnetic field could reach a few hundred G or even a few kG \citep{Saar}. Spectropolarimetry has been used to infer a surface magnetic field strength for many nearby M dwarfs using Zeeman Doppler Imaging \citep[e.g.][]{moutou2016, Moutou2017}. Magnetic fields of hundreds of gauss would make the stellar ECM emission detectable with current radio observatories such as LOFAR \citep{LOFAR} and FAST \citep{FAST}. We encourage follow-up observation of TOI-6255 using these radio telescopes.

\subsection{Is TOI-6255 b already disintegrating?}

Given TOI-6255 b's proximity to the Roche limit (see Section \ref{sec:discussion_distortion}), one may wonder whether TOI-6255 b is already losing mass, making it similar to the ``disintegrating planets'' KOI-2700 b \citep{Rappaport2700}, KIC 12557548 b \citep{Rappaport1255}, and K2-22b \citep{SanchisK2-22}. All three of those disintegrating planets have longer orbital periods than TOI-6255 b (Fig. \ref{fig:tidal}). The transits of all three disintegrating planets are notably asymmetric ( egress can be several times longer than ingress) about the mid-transit times, and there can be an order of unity change in the transit depths over time. The prevailing explanation is that we are observing a tail of dusty materials episodically emanating from these planets \citep[]{vanLieshout}. See also upcoming JWST observations of K2-22 b by Tusay et al. (Program 3315). In addition, due to their material strength, 
up-to-km-size and less conspicuous debris from an ongoing 
disrupting planet may be preserved despite the intense tidal
tearing in the proximity of the host star's surface\citep{zhangyun2021}. 

We show that TOI-6255 b is likely not undergoing catastrophic disintegration. 
We did not detect any asymmetry in the {\it TESS} transit signal. We fitted a simple trapezoidal model to the phase-folded light curve similar to \citet{SanchisK2-22}, and found no statistical significant difference between the ingress and egress of the light curve. Moreover, the transit depth of TOI-6255 b also appears to be constant. The standard deviation of in-transit and out-of-transit flux residuals of each {\it TESS} 2-min exposures are almost identical (1415 ppm v.s. 1429 ppm). Moreover, the transit depths in {\it TESS} Sectors 16 and 56 (separated by three years) are consistent within 1 $\sigma$ of each other. \citet{Perez-Becker} showed that catastrophic evaporation only takes place on small (0.1 $M_\oplus$), strongly irradiated (2000 K) planets (i.e. high thermal sound speed and low escape velocity from the planet). TOI-6255 b (1.44 $M_\oplus$) is likely too massive to undergo this mode of mass loss, although with its $P_{\rm orb}/P_{\rm Roche} \simeq 1.13$, 
Roche-lobe overflow of the planet's atmosphere may significantly ease this stringent requirement\citep{lishulin2010}.  

\subsection{Why the shortest-period planets orbit late-type stars?}

An intriguing observation is that TOI-6255 b and other top contenders for the shortest period planet -- KOI-1843.03 \citep[4.2-hours, M1 host][]{Ofir2013,Rappaport}, K2-137 b \citep[4.3-hours, M3 host][]{Smith}, K2-141 b \citep[6.7-hour, K7 hosts][]{Barragan141}, GJ-367 b \citep[7.7-hours, M1 host][]{Lam} -- are all around late-type stars. The disintegrating planets are also all around K or M-type host stars. \citet{USP} showed that the occurrence rate for USPs is about 0.15$\pm$0.05 \% for F stars, 0.51$\pm0.07$\% for G stars, but increases to 1.1$\pm$0.4\% for M dwarfs. Why are the shortest period planets found around late-type stars? What would be the implication for planet formation at the inner edge of the protoplanetary disks?

\citet{Lee_usp} noted that the inner edge of the protoplanetary disk plays a pivotal role in setting the lower boundary for the orbital periods of planets. Magnetospheric disk truncation is expected to happen at the corotation radius set by the rotation period of the host star. Rotation measurements of young clusters \citep[see e.g. Fig. 2 of][]{Bouvier} have indeed indicated that late-type stars tend to maintain fast rotation throughout the timescale of planet formation (with periods often $<3$ days on 10 Myr timescales). On the other hand, sun-like stars tend to spin down faster even before the planetary disk dissipates. It could be the case that the smaller disk inner edge around late-type stars enables the formation of more short-period planets like TOI-6255 b before tides further shrink their orbits over Gyr timescale.

\subsection{JWST Phase Curve Observations}
\label{sec:discussion_phase_curve}
USP planets like TOI-6255 b are particularly amenable to phase curve and secondary eclipse observations with the James Webb Space Telescope \citep[JWST;][]{Gardner}. The flux ratio between USPs and their host stars is at a favorable level of a few hundred parts per million in the infrared. Furthermore, the USPs are expected to be fully tidally locked with their host star  \citep{Winn2018}, resulting in a constant day side and a constant night side. Phase curve observations map out the temperature distribution as a function of longitudes on the planet \citep{Koll}, which in turn provides insights into the presence or absence of an outgassed, high-molecular-weight atmosphere \citep{Demory}.  In the absence of an atmosphere, the same measurements may directly probe the surface mineralogy on these planets \citep{Hu2012,Kreidberg,Zhang2024}. The stellar insolation TOI-6255 b receives is about 590$\pm90F_\oplus$; the equilibrium temperature is 1340$\pm60$ K assuming a low albedo of 0.1 appropriate for dark basaltic surface \citep{Essack}. The dayside may have a different extent of lava pool, depending on the detail of heat transport and if there is substantial tidal heating \citep{kite2016}. 

Ten USP planets (GJ 367 b, K2-141 b, 55 Cnc e, LHS 3844 b, TOI-561 b, WASP-47 e, TOI-1685 b, TOI-4481 b, and TOI-2445 b) have been selected as JWST targets. The majority of these are super-Earths (>1.2 $R_\oplus$, Fig. \ref{fig:esm}). This preference for super-Earths is mainly driven by signal-to-noise considerations, as the Emission Spectroscopy Metric \citep[ESM;][]{Kempton} is proportional to $(R_p/R_\star)^2$. TOI-6255, on the other hand, is an Earth-sized planet (1.079$\pm$0.065 $R_\oplus$) on a 5.7-hour orbit around a nearby M dwarf ($K=8.1$). It boasts an ESM of 24, making it a top ESM rocky planet for phase curve observations with MIRI/LRS, trailing only LHS 3844 b \citep{Vanderspek} and 55 Cnc e \citep{Winn2011,Crida}.

Finally, we examined how the tidal distortion of a planet can modify the phase curve variations. The obvious effect is the ellipsoidal light variation (ELV). Since TOI-6255 b is likely in synchronous rotation, the long axis of the ellipsoidal planet always points toward the star. Hence, at different orbital phases, we should see a different projection of the planet as described in Eqn \ref{eqn:projected_area}, which roughly scales as ($\sqrt{\sin^2(\phi)}$; $\phi$ is the orbital phase). The projection effect has two peaks at both quadratures of the orbit. To first order, phase curve variations of the planet will be modulated by this area projection effect. This effect is analogous to the ellipsoidal light variation of the binary stars \citep[e.g.][]{Morris}. For TOI-6255, we calculated that the stellar ELV is $<1$ ppm, whereas the planetary ELV is tens of ppm in amplitude depending on the tidal Love number and the strength of thermal emission. The planetary ELV is directly proportional to the change of the projected area of the tidal distorted planet (i.e. 10\% fractional area change depending on the tidal Love number $h_2$).  In principle, the planetary ELV can be distinguished from the thermal phase curve.  The thermal emission of a bare rocky planet only peaks at the secondary eclipse, whereas the ELV peaks at quadratures ($-\cos(\phi)$ v.s. $\sqrt{\sin^2(\phi)}$). The tidal distortion of TOI-6255 may further modify the phase curve variation through geological effects. The long axis of the planet (i.e. at the substellar and antistellar points) is where the silicate mantle is the thickest, and the thicker mantle there may promote convection and volcanic activity. This effect would result in an even hotter sub-stellar point. However, a proper simulation of this effect and their observability with JWST requires a dedicated model; we defer that to future work.

\section{Summary}
In this work, we present the discovery and confirmation of TOI-6255 b using transit observations and Doppler monitoring of the host star. We summarize the key findings as follows:

\begin{itemize}
    \item {TOI-6255 b has an orbital period of 5.7 hours and a transit radius of 1.079$\pm$0.065 $R_\oplus$.}

      \item {By applying both Gaussian Process regression and the Floating Chunk Offset method on the KPF and CARMENES radial velocity measurements, we constrained the mass of TOI-6255 b to be 1.44$\pm$0.14 $M_\oplus$.}

      \item {TOI-6255 b has iron core mass fraction of 45$\pm$32\% \citep[or 41$\pm$20\% if using the model by \citealt{Plotnykov2020}]{Zeng2016}, which is broadly consistent with an Earth-like composition.}

      \item {TOI-6255 b experiences extreme tidal forces. It is marginally outside the Roche Limit with $P_{\rm orb}/P_{\rm Roche}$ = 1.13 $\pm0.10$ \citep{Rappaport}. The planet would tidally decay to the Roche Limit within the next 400 Myr if the host star's reduced tidal quality factor is $Q_\star^\prime=10^7$. The resultant tidal disruption may produce the long-suspected planet engulfment signature.}
      
              \item {TOI-6255 b may experience star-planet magnetic interaction analogous to our Io-Jupiter interaction. Resultant radio emission may be orbital-phase-modulated and circularly polarized.}
      
      \item {TOI-6255 b is likely ellipsoidal in shape. Assuming a tidal Love number $h_2=1$, the planet's long axis is about 10\% longer than its short axis. Accounting for this effect, the CMF of the planet is reduced to 31$\pm$30\%. The ellipsoidal shape produces a subtle distortion of the transit light curve ($<10$ ppm) that could not be detected with the existing {\it TESS} data.}

       \item {With an ESM of 24, TOI-6255 b is a top-ranking target for JWST phase curve variations which may determine the presence/lack of a heavy-mean-molecular-weight atmosphere. If the planet is a bare rocky core, emission spectra may tell us the dominant surface mineralogy. The ellipsoidal shape of TOI-6255 further modifies the phase curve variation; the planetary ellipsoidal light variation should enhance planetary emission near quadratures of the orbit.}

\end{itemize}

\software{{\sc AstroImage} \citep{Collins:2017}, {\sc Isoclassify} \citep{Huber},{\sc isochrones} \citep{Morton} {\sc MIST} \citep{MIST}, {\sc SpecMatch-Syn} \citep{Petigura_thesis}, {\sc Batman} \citep{Kreidberg2015}, {\sc emcee} \citep{emcee}}

\facilities{Keck I: (KPF), 3.5\,m Calar Alto, {\it TESS}, LCOGT, MuSCAT2, MuSCAT3, WASP, Palomar}

\begin{acknowledgments}
\begin{center}
ACKNOWLEDGEMENTS
\end{center}

We thank Henrique Reggiani for an independent investigation of the stellar spectrum. We thank Doug Lin, Luke Bouma, Eugene Chiang, Darryl Seligman, Saul Rappaport, Jennifer van Saders, Ji Wang, and Li Zeng for insightful discussions. We also thank Connie Rockosi for their contribution to successful construction of KPF. 

Support for this work was provided by NASA through the NASA Hubble Fellowship grant HST-HF2-51503.001-A awarded by the Space Telescope Science Institute, which is operated by the Association of Universities for Research in Astronomy, Incorporated, under NASA contract NAS5-26555.

A NASA Key Strategic Mission Support titled ``Pinning Down Masses of JWST Ultra-short-period Planets with KPF'' (PI: F. Dai) provided the telescope access and funding for the completion of this project.

This work was supported by a NASA Keck PI Data Award, administered by the NASA Exoplanet Science Institute. Data presented herein were obtained at the W. M. Keck Observatory from telescope time allocated to the National Aeronautics and Space Administration through the agency's scientific partnership with the California Institute of Technology and the University of California. The Observatory was made possible by the generous financial support of the W. M. Keck Foundation.

The data presented herein were obtained at the W. M. Keck Observatory, which is operated as a scientific partnership among the California Institute of Technology, the University of California and the National Aeronautics and Space Administration. The Observatory was made possible by the generous financial support of the W. M. Keck Foundation.

The authors wish to recognize and acknowledge the very significant cultural role and reverence that the summit of Maunakea has always had within the indigenous Hawaiian community.  We are most fortunate to have the opportunity to conduct observations from this mountain.

We acknowledge the use of public TESS data from pipelines at the TESS Science Office and at the TESS Science Processing Operations Center. Resources supporting this work were provided by the NASA High-End Computing (HEC) Program through the NASA Advanced Supercomputing (NAS) Division at Ames Research Center for the production of the SPOC data products.

This paper made use of data collected by the TESS mission and are publicly available from the Mikulski Archive for Space Telescopes (MAST) operated by the Space Telescope Science Institute (STScI). 
 
Funding for the TESS mission is provided by NASA’s Science Mission Directorate. KAC and CNW acknowledge support from the TESS mission via subaward s3449 from MIT.

This research has made use of the Exoplanet Follow-up Observation Program (ExoFOP; DOI: 10.26134/ExoFOP5) website, which is operated by the California Institute of Technology, under contract with the National Aeronautics and Space Administration under the Exoplanet Exploration Program.

This work makes use of observations from the LCOGT network. Part of the LCOGT telescope time was granted by NOIRLab through the Mid-Scale Innovations Program (MSIP). MSIP is funded by NSF. 

This work is partly supported by JSPS KAKENHI Grant Number JPJP24H00017 and JSPS Bilateral Program Number JPJSBP120249910.
This article is based on observations made with the MuSCAT2 instrument, developed by ABC, at Telescopio Carlos S\'anchez operated on the island of Tenerife by the IAC in the Spanish Observatorio del Teide.
This paper is based on observations made with the MuSCAT3 instrument, developed by the Astrobiology Center and under financial supports by JSPS KAKENHI (JP18H05439) and JST PRESTO (JPMJPR1775), at Faulkes Telescope North on Maui, HI, operated by the Las Cumbres Observatory.

This research was carried out, in part, at the Jet Propulsion Laboratory and the California Institute of Technology under a contract with the National Aeronautics and Space Administration and funded through the President’s and Director’s Research \& Development Fund Program.

We acknowledge financial support from the Agencia Estatal de Investigaci\'on of the Ministerio de Ciencia e Innovaci\'on MCIN/AEI/10.13039/501100011033 and the ERDF “A way of making Europe” through project PID2021-125627OB-C32, and from the Centre of Excellence “Severo Ochoa” award to the Instituto de Astrofisica de Canarias.

D.H. acknowledges support from the Alfred P. Sloan Foundation, the National Aeronautics and Space Administration (80NSSC21K0652), and the Australian Research Council (FT200100871).

DRC and CAC acknowledge support from NASA 18\-2XRP18\_2\-0007

\end{acknowledgments}

\bibliography{main}

\appendix

\counterwithin{figure}{section}

\section{Supplementary Figures}

\begin{figure}
\center
\includegraphics[width = .5\columnwidth]{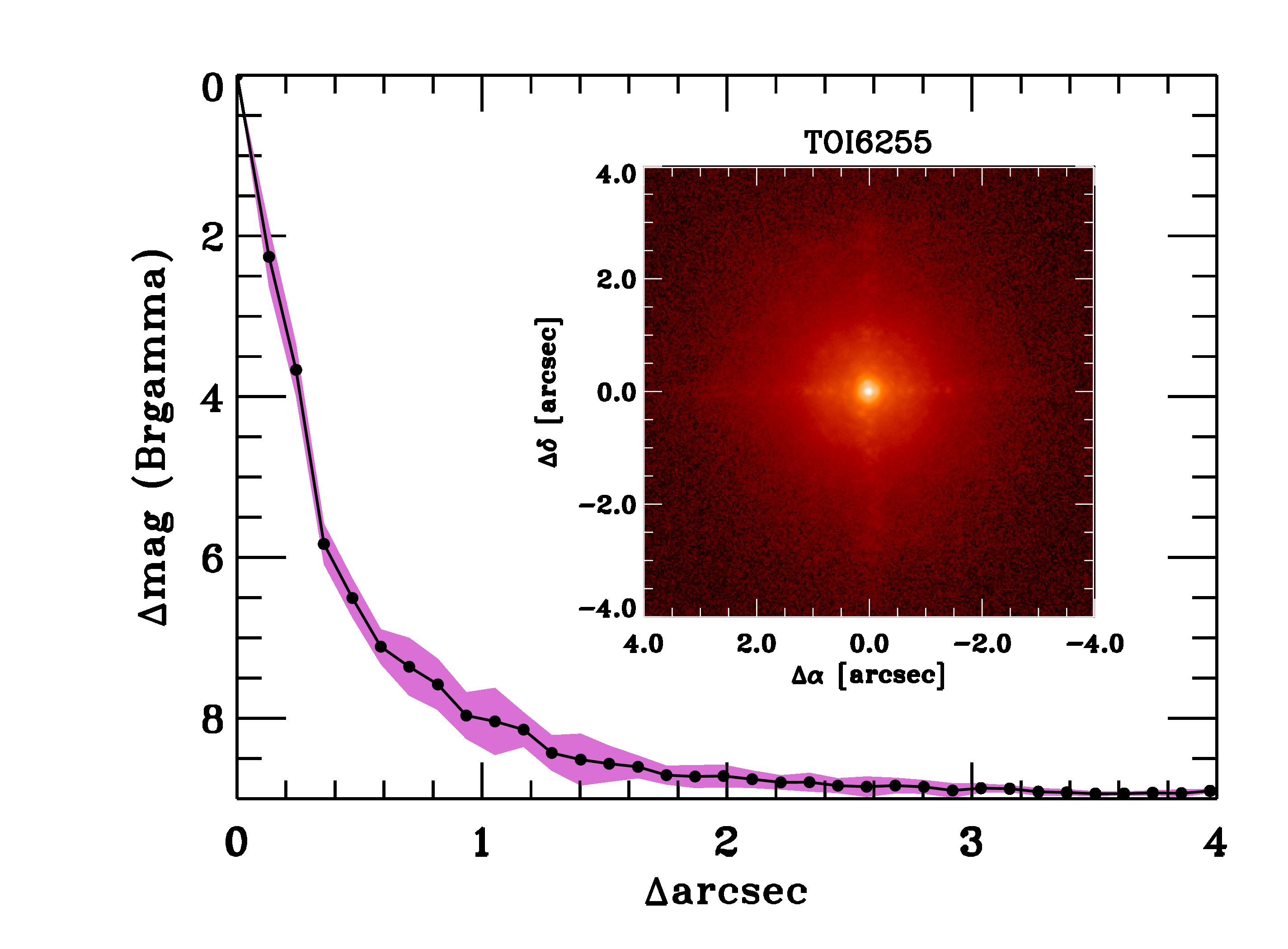}
\caption{The contrast curve as function of radial distance from TOI-6255 using Palomar/PHARO observations \citep{hayward2001}. The inset shows the high contrast image of TOI-6255. We could rule out companions <6.5 mag within at 0.5''.}
\label{fig:palomar}
\end{figure}

\begin{figure}

\center
\includegraphics[width = .52\columnwidth]{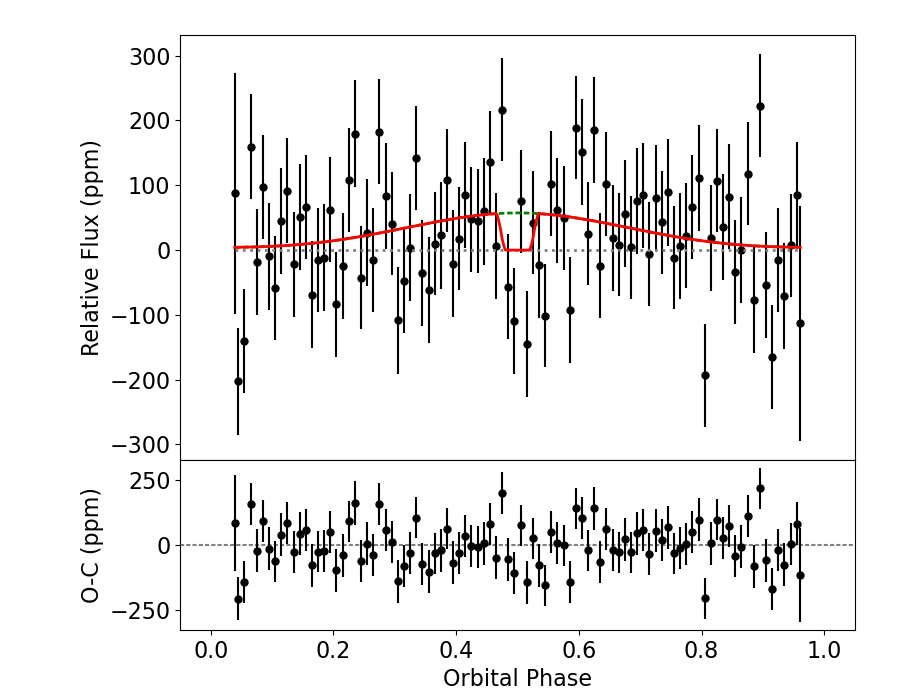}
\includegraphics[width = .5\columnwidth]{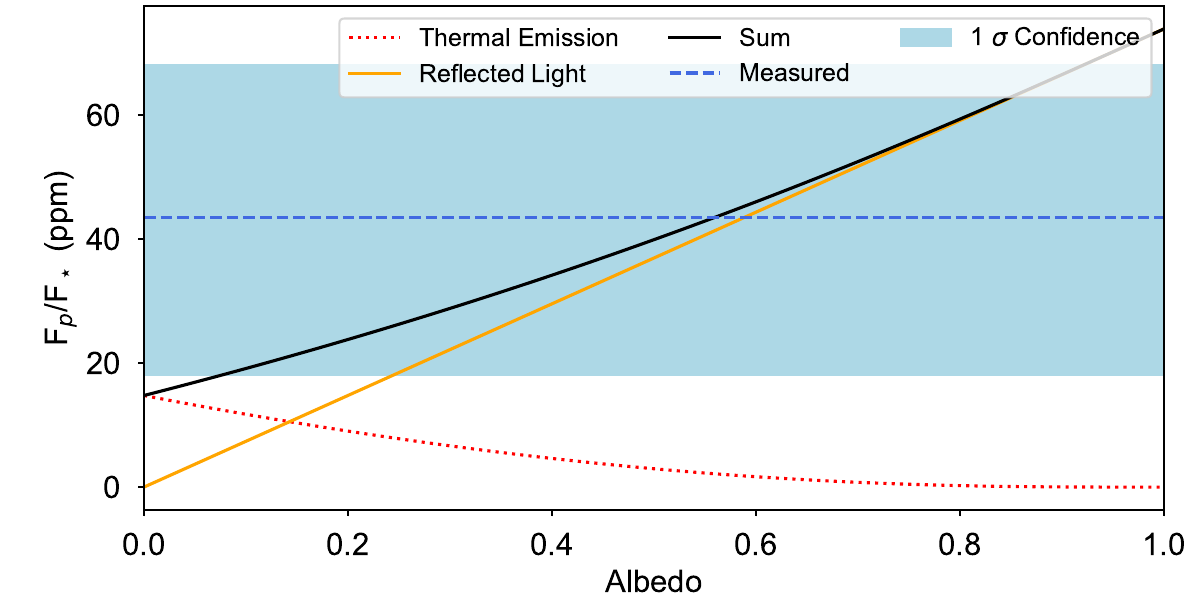}
\caption{{\bf Top:} The phase curve variation of TOI-6255 b observed by {\it TESS}.  The red curve shows the best-fit phase curve and secondary eclipse model. The phase curve is only detected at a 2-$\sigma$ level with two sectors of {\it TESS} observations in the optical. {\bf Bottom:} The amount of thermal emission (red dotted line) and reflected light (orange solid line) in the {\it TESS} band as a function of the Bond Albedo. The blue dashed line is the measured amount of planetary flux over the stellar flux ($F_p/F_\star$) and the associated 1-$\sigma$ confidence interval (blue area). The phase curve variation in {\it TESS} is likely a combination of thermal emission and reflected light.}
\label{fig:tess_phase_curve}
\end{figure}

\begin{figure*}\label{fig:periodograms}
\center
\includegraphics[width = 1.\columnwidth]{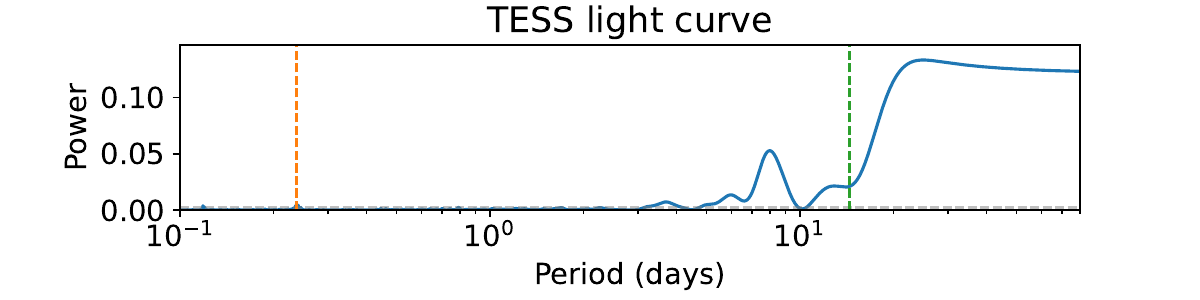}
\includegraphics[width = 1.\columnwidth]{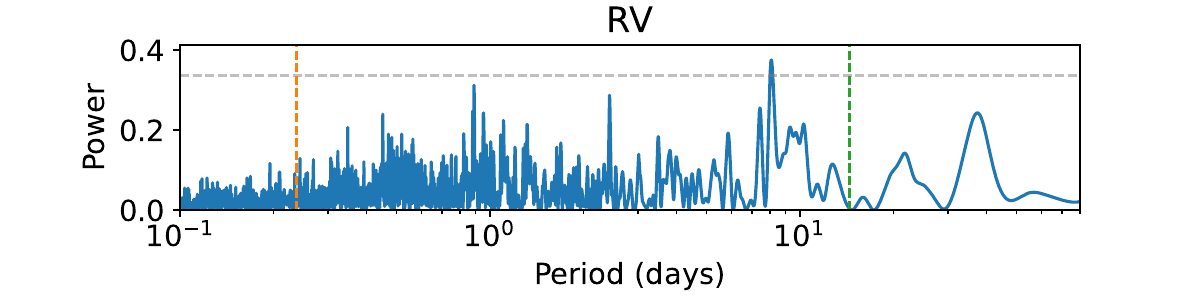}
\includegraphics[width = 1.\columnwidth]{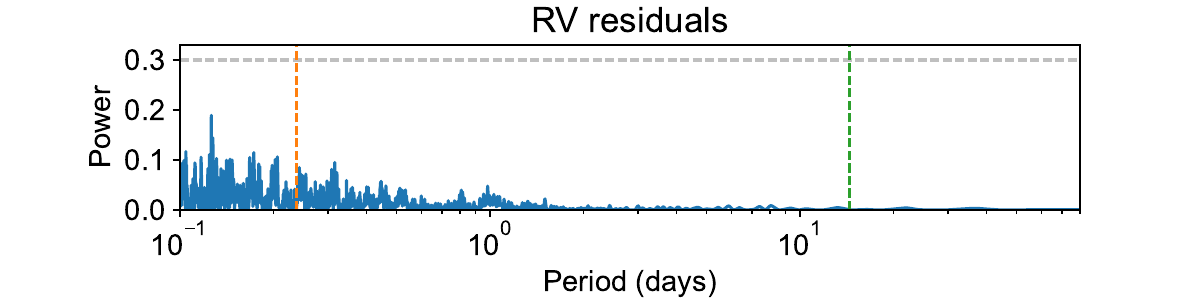}
\includegraphics[width = 1.\columnwidth]{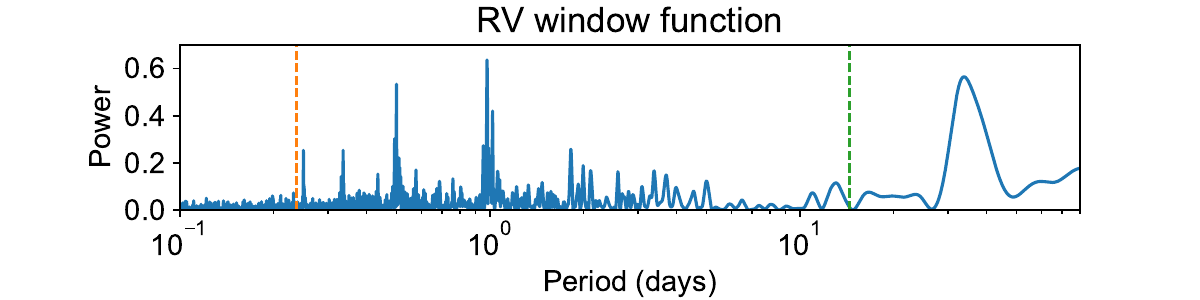}
\caption{The Lomb-Scargle periodograms of the {\it TESS} light curve of TOI-6255, the RV variations, the RV residuals after removing the best-fit model, and the RV window function (RV sampling). The vertical dotted lines show the orbital periods of TOI-6255 b and TOI-6255.02. The planetary radial velocity signal is masked by a combination of stellar activity and instrumental variation. The horizontal dotted line indicates 0.1\% false positive level.}
\end{figure*}

\begin{figure}
\center 
\includegraphics[width = .5\columnwidth]{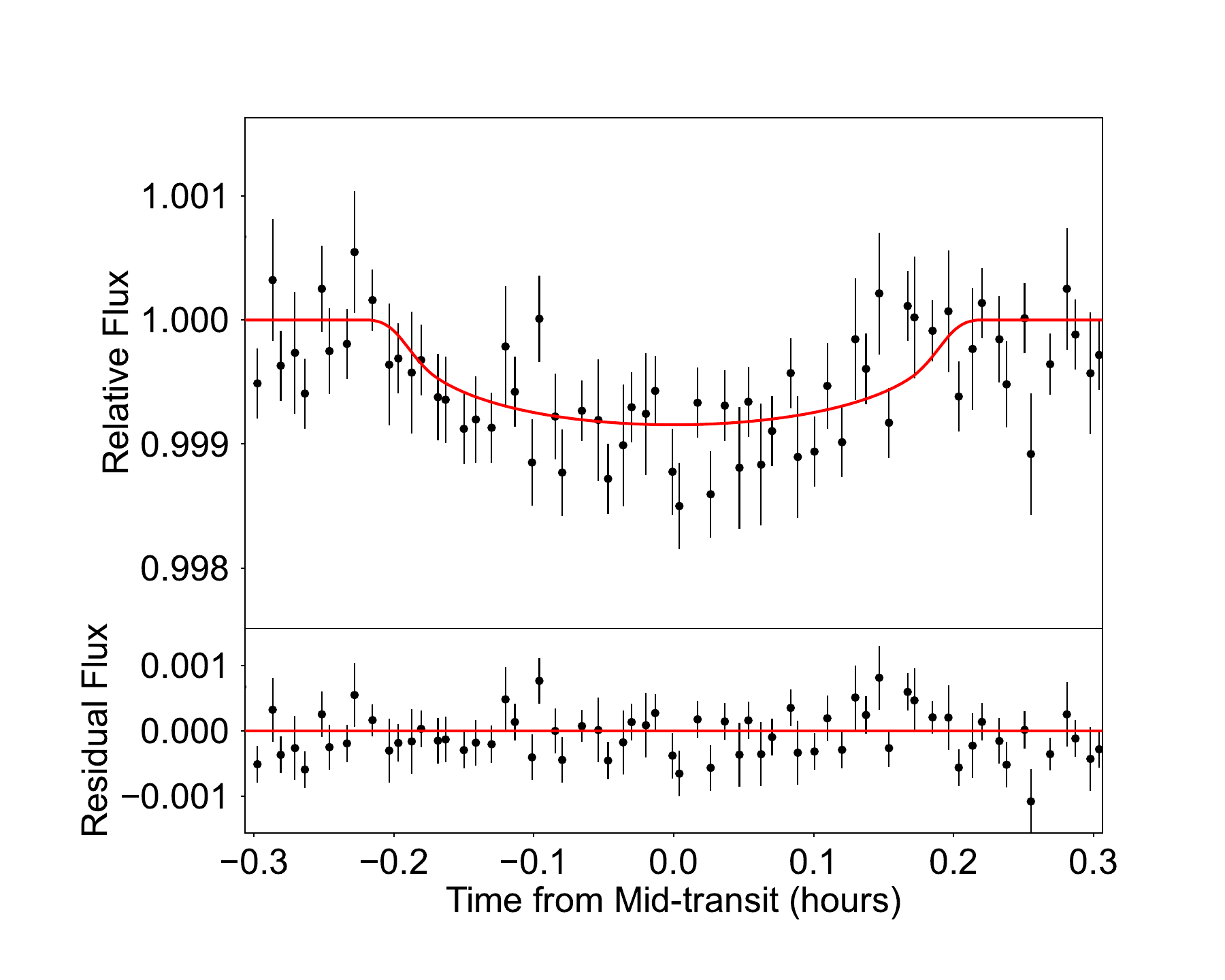}
\caption{Phase-folded and binned MuSCAT3 \citep{Narita_muscat3} observation of TOI-6255 b. The transit parameters inferred from MuSCAT3 observations are consistent with those from {\it TESS}.}
\label{fig:transit_muscat3}
\end{figure}

\begin{deluxetable}{ccccc}
\tabletypesize{\scriptsize}
\tablecaption{Keck/KPF Radial Velocities of TOI-6255}
\label{tab:rv}
\tablehead{
\colhead{Time (BJD)} & \colhead{RV (m/s)} & \colhead{RV Unc. (m/s)} }
\startdata
2460087.06212686 & 7.91 & 0.95\\
2460090.08283583 & 2.73 & 0.91\\
2460091.04582747 & -5.73 & 0.84\\
2460093.05386041 & -11.03 & 1.06\\
2460093.11093392 & -12.58 & 1.05\\
2460100.02074132 & -10.36 & 1.00\\
2460100.07705751 & -4.79 & 1.12\\
2460100.98823238 & -2.14 & 0.95\\
2460101.02651958 & -0.00 & 0.87\\
2460106.06587267 & -16.29 & 0.89\\
2460106.10595727 & -18.94 & 0.91\\
2460124.04069331 & -11.19 & 0.89\\
2460124.11356605 & -6.04 & 0.96\\
2460126.07766735 & 9.79 & 1.03\\
2460126.11463478 & 6.95 & 0.89\\
2460129.99777854 & -9.10 & 0.97\\
2460130.99139985 & -5.80 & 1.00\\
2460135.10930206 & -0.66 & 1.01\\
2460136.98291478 & -8.14 & 1.49\\
2460137.07709318 & -11.15 & 1.21\\
2460146.02819387 & -1.36 & 1.02\\
2460146.10394072 & -3.63 & 0.98\\
2460150.02103948 & 15.30 & 0.86\\
2460150.09242876 & 21.43 & 1.00\\
2460153.01666907 & 13.54 & 1.06\\
2460153.06132063 & 10.06 & 0.93\\
2460154.90337595 & 14.78 & 1.00\\
2460155.91078596 & 3.31 & 0.94\\
2460157.02830366 & -1.22 & 0.96\\
2460157.05650057 & -2.94 & 0.94\\
2460157.06840406 & -2.94 & 0.98\\
2460185.76215501 & -7.86 & 1.41\\
2460185.91568898 & -8.41 & 1.46\\
2460186.05100537 & -6.90 & 1.41\\
2460187.84396189 & -10.25 & 1.01\\
2460188.00378558 & -2.40 & 1.01\\
2460188.10885611 & -7.88 & 1.29\\
2460188.77511581 & -6.39 & 0.93\\
2460188.86314294 & -1.67 & 0.96\\
2460188.96253992 & 2.61 & 1.08\\
2460190.00555127 & 1.58 & 0.92\\
2460190.04392353 & 4.74 & 1.00\\
2460190.07231719 & 7.42 & 1.04\\
2460191.00661120 & 5.15 & 1.00\\
2460191.05434485 & 10.08 & 0.92\\
2460191.07363138 & 14.80 & 0.81\\
\enddata
\end{deluxetable}

\begin{deluxetable}{ccccc}
\tabletypesize{\scriptsize}
\tablecaption{Tab. 3 Continued}
\label{tab:rv2}
\tablehead{
\colhead{Time (BJD)} & \colhead{RV (m/s)} & \colhead{RV Unc. (m/s)} }
\startdata
2460192.02226241 & 15.08 & 0.92\\
2460192.03837202 & 14.70 & 0.92\\
2460192.09976076 & 14.96 & 0.97\\
2460192.76651900 & 3.67 & 1.11\\
2460193.77734790 & 3.34 & 1.35\\
2460193.82599660 & -1.97 & 1.31\\
2460195.76304918 & -5.14 & 0.94\\
2460195.80161774 & -2.75 & 1.05\\
2460195.83124630 & -2.24 & 0.97\\
2460196.76153753 & -7.40 & 1.02\\
2460196.80613649 & -3.91 & 1.02\\
2460196.85435330 & -4.87 & 0.94\\
2460198.74800479 & 11.49 & 0.91\\
2460198.81742941 & 7.85 & 0.87\\
2460198.84542413 & 5.39 & 0.77\\
2460200.74046532 & -7.21 & 1.07\\
2460200.78534718 & -2.20 & 1.08\\
2460200.83814583 & -1.26 & 1.07\\
2460201.77332723 & 6.47 & 0.73\\
2460201.79622377 & 9.70 & 0.81\\
2460201.85800751 & 7.00 & 0.81\\
2460204.77482518 & -21.49 & 0.96\\
2460204.80641455 & -20.91 & 0.88\\
2460204.91623978 & -15.53 & 0.99\\
2460272.70896174 & 8.38 & 0.92\\
2460272.71643373 & 11.46 & 0.98\\
2460272.72410032 & 11.83 & 1.07\\
2460272.73387913 & 10.94 & 0.96\\
2460272.74149438 & 14.75 & 0.98\\
2460272.74894656 & 14.15 & 1.10\\
2460272.75645266 & 13.54 & 1.01\\
2460272.76405543 & 14.87 & 0.98\\
2460272.77146772 & 14.92 & 0.92\\
2460272.78168415 & 13.15 & 0.95\\
2460272.78911734 & 16.08 & 1.07\\
2460272.79657609 & 15.12 & 1.10\\
2460272.80441531 & 13.52 & 0.97\\
2460272.81143582 & 13.67 & 1.02\\
2460272.81907260 & 13.19 & 1.03\\
2460272.82888544 & 11.59 & 1.04\\
2460272.83632523 & 13.13 & 1.07\\
2460272.84397265 & 10.70 & 1.14\\
2460272.85134004 & 11.59 & 1.10\\
2460272.85912790 & 8.78 & 1.12\\
2460272.86647520 & 10.47 & 1.14\\
\enddata
\end{deluxetable}

\begin{deluxetable}{ccccc}
\tabletypesize{\scriptsize}
\tablecaption{CARMENES Radial Velocities of TOI-6255}
\label{tab:rv3}
\tablehead{
\colhead{Time (BJD)} & \colhead{RV (m/s)} & \colhead{RV Unc. (m/s)} }
\startdata
 2460120.46393455 & -3.06 & 1.62\\
 2460120.51080524 & -0.05 & 1.52\\
 2460120.58599291 & 2.33 & 1.33\\
 2460120.64040814 & 0.25 & 1.31\\
 2460121.47968095 & 2.55 & 3.92\\
 2460122.46464087 & 11.23 & 4.30\\
 2460122.52944385 & -0.20 & 2.53\\
 2460122.56474435 & 2.99 & 2.34\\
 2460122.60303551 & -0.93 & 1.97\\
 2460123.46324037 & 6.29 & 3.24\\
 2460123.52151940 & -3.84 & 1.94\\
 2460123.57083056 & -3.55 & 1.81\\
 2460123.61803215 & -3.06 & 1.74\\
 2460124.46376153 & -4.43 & 2.35\\
 2460124.52255125 & -5.01 & 1.82\\
 2460124.57117621 & -5.00 & 2.85\\
 2460124.61768596 & -0.29 & 1.90\\
 2460126.46060562 & 0.59 & 3.39\\
 2460126.49574863 & 4.84 & 2.23\\
 2460126.56197642 & 3.09 & 1.83\\
 2460126.60298822 & 5.36 & 1.55\\
 2460160.37799236 & 5.27 & 3.90\\
 2460160.42652334 & 1.76 & 2.57\\
 2460160.48450259 & 1.42 & 3.46\\
 2460160.52696557 & 3.37 & 3.23\\
 2460160.57725878 & 5.95 & 2.67\\
 2460160.62478288 & 3.84 & 2.38\\
 2460164.37810974 & -1.80 & 2.52\\
 2460164.42647410 & 0.01 & 1.81\\
 2460164.48147214 & -1.44 & 2.08\\
 2460164.53039479 & -2.59 & 2.06\\
 2460164.57831531 & -4.88 & 2.46\\
 2460164.63229627 & -2.66 & 2.30\\
\enddata
\end{deluxetable}

\section{Phase Curve and Secondary Eclipse}\label{sec:app_phase_curve}

Given the extremely short orbital period of TOI-6255 b, we searched for any phase curve variation or secondary eclipse in the {\it TESS} light curve. We removed data taken during the transits of TOI-6255 b. We then detrended any long-term  stellar variability and instrumental effects following the method of \citet{Sanchis78}. Briefly, for each data point in the original light curve, we isolated a window of 1$\times$, 2$\times$ or 3$\times P_{\rm orb}$ the orbital period around that data point. We removed any 5-$\sigma$ outliers, and then fitted a linear function of time to the remaining data points within that window. The original data point is then divided by the local best fit linear function of time. We experimented with both the PDC-SAP and SAP {\it TESS} light curves as well as using different window width (1$\times$, 2$\times$ or 3$\times P_{\rm orb}$). The resultant phase curve variations did not show any substantial differences between these choices. We report results based on the SAP light curve and with window width of 2$P_{\rm orb}$).

Our occultation (secondary eclipse) model uses the best-fit system parameters (period, orbital inclination, etc.) from the much more precise transit (primary eclipse) model (Section \ref{sec:transit_model}). The only difference is that the limb darkening coefficients have been set to 0, and the secondary eclipse depth ($\delta_{\rm sec}$) is allowed to vary freely to account for a combination of reflected stellar light and thermal emission from the planet in the {\it TESS} band (600-1000 nm). The out-of-eclipse phase curve variation is modeled as a Lambertian disk \citep[see e.g.][]{Demory}. This model is characterized by the amplitude of illumination effect $A$ and any phase offset of the peak of the phase curve $\theta$. We initially let $\delta_{\rm sec}$ and $A$ to vary independently, in essence allowing for a non-zero flux from the night side of the planet. However, TOI-6255 has only been observed by {\it TESS} for two sectors, the SNR of the existing data does not support a statistically significant detection of the nightside flux or a phase offset (see Fig. \ref{fig:tess_phase_curve}). We therefore report the results of the simplest phase curve model where: $\delta_{\rm sec}\equiv$$A$ (no night-side contribution) and  $\theta=0$ (no phase offset). We sampled the posterior distribution using another MCMC analysis with {\tt emcee} \citep{emcee}. The procedure is similar to that described in the previous section. We found that $\delta_{\rm sec}\equiv$$A_{\rm ill} = 43\pm25$ ppm. This is less than 2-$\sigma$ detection of the phase curve/secondary eclipse.

In Fig. \ref{fig:tess_phase_curve}, we plot the ratio of planetary to stellar flux $F_p/F_\star$ in the {\it TESS} band as a function of the planet's Bond Albedo. We assumed that the albedo in the TESS band is the same as the Bond Albedo. The higher the Bond albedo, the more the reflective the planet is, and the lower the equilibrium temperature will be. The planet reflects more stellar light but gives out weaker thermal emission. The blue-shaded area is the 1-$\sigma$ confidence interval from {\it TESS}. The confidence interval is so wide that we cannot place a meaningful constraint on the planet's albedo. TOI-6255 b is a top-ranking target for future JWST phase curve characterization in the near and mid-infrared, any phase curve variation of TOI-6255 will be much more observable in the infrared (see Section \ref{sec:discussion_phase_curve}).

\section{Transit Modeling with Other Light Curves}\label{sec:transits}
TOI-6255 b was observed by a suite of ground-based facilities including the Las Cumbres Observatory Global Telescope \citep[LCOGT;][]{Brown:2013} network nodes at the Teide Observatory (1-m telescope), the McDonald Observatory (1-m telescope), MuSCAT2 \citep[1.5-m,][]{Narita_muscat2}, and LCOGT MuSCAT3 \citep[2-m,][]{Narita_muscat3}. 

Due to the small size of the transit signal ($\sim$0.7 mmag), only MuSCAT3 (the largest aperture) detected the transit signal robustly. We highlight the results from MuSCAT3 here. All other ground-based photometric observations can be downloaded from the ExoFOP website. MuSCAT3 is mounted on the 2-m Faulkes Telescope North at Haleakala Observatory on Maui, Hawai'i. MuSCAT3 has 4 simultaneous channels: $g^\prime$ (400-550 nm), $r^\prime$ (550-700 nm), $i^\prime$ (700-820 nm), and $z_s$ (820-920 nm). We combined the four channels together in our photometric analysis using their weighted mean and a single set of limb darkening coefficients. 

TOI-6255 was observed by MuSCAT3 on UT Aug 14th, Aug 18th, Sept 8th, Sept 28th, and Nov 24th of 2023.  The images  were calibrated using standard procedures by the BANZAI pipeline \footnote{\url{https://github.com/LCOGT/banzai}}. Differential photometric data were extracted using {\tt AstroImageJ} \citep{Collins:2017} with circular photometric apertures having radius $6\arcsec$ or smaller that excluded flux from the nearest known neighbor (Gaia DR3 1956328333129802112) that is $9\arcsec$ west of TOI-6255. We followed the same procedure as our analysis of the {\it TESS} light curves. The phase-folded and binned MuSCAT3 light curve and the best-fit transit model are shown in Fig. \ref{fig:transit_muscat3}. The posterior distribution of the transit parameters are consistent with that from {\it TESS} albeit with higher uncertainty. Notably, the MuSCAT3 observations also suggest a high impact parameter (b=0.77$\pm$0.09) for TOI-6255 b (c.f. b = 0.84$\pm0.05$ from {\it TESS}). MuSCAT3 suggests  $R_p/R_\star$ = 0.0286$\pm$0.0031 which agrees with the  $R_p/R_\star$ = 0.0267$\pm$0.0014 from {\it TESS}) well within 1-$\sigma$.  

We also modeled the TESS Gaia Light Curve \citep[TGLC, ][]{Han} which used point-spread function (PSF) modeling to remove contamination from nearby stars in the {\it TESS} 200-sec Full-Frame-Image light curves. TGLC light curve is only available for Sector 56 for TOI-6255. We found that SPOC/PDC-SAP and TGLC produced consistent results on the radius of TOI-6255 b. TGLC \citep{Han} light curves also give a consistent $R_p/R_\star$ = 0.0273$\pm$0.0014.

\end{document}